\documentclass[a4paper,12pt, epsfig]{article}
\usepackage{epsfig}
\usepackage{amssymb,latexsym}
\usepackage{amsfonts}
\usepackage{amsmath}
\usepackage[colorlinks, linkcolor=blue, citecolor=blue, urlcolor=blue]{hyperref}
\usepackage{epstopdf}
\usepackage{graphicx}
\pdfoutput=1

\newskip\humongous \humongous=0pt plus 1000pt minus 1000pt

\newif\ifdtup

\catcode`\@=11

\@addtoreset{equation}{section}
\def\theequation{\thesection.\arabic{equation}}

\def\@normalsize{\@setsize\normalsize{15pt}\xiipt\@xiipt
\abovedisplayskip 14pt plus3pt minus3pt%
\belowdisplayskip \abovedisplayskip
\abovedisplayshortskip \z@ plus3pt%
\belowdisplayshortskip 7pt plus3.5pt minus0pt}

\def\small{\@setsize\small{13.6pt}\xipt\@xipt
\abovedisplayskip 13pt plus3pt minus3pt%
\belowdisplayskip \abovedisplayskip
\abovedisplayshortskip \z@ plus3pt%
\belowdisplayshortskip 7pt plus3.5pt minus0pt
\def\@listi{\parsep 4.5pt plus 2pt minus 1pt
      \itemsep \parsep
      \topsep 9pt plus 3pt minus 3pt}}

\relax

\catcode`@=12

\catcode`\@=11

\def\section{\@startsection{section}{1}{\z@}{3.5ex plus 1ex minus
    .2ex}{2.3ex plus .2ex}{\large\bf}}

\def\thesection{\arabic{section}}
\def\thesubsection{\arabic{section}.\arabic{subsection}}

\def\appendix{\setcounter{section}{0}
  \def\thesection{Appendix \Alph{section}}
  \def\thesubsection{\Alph{section}.\arabic{subsection}}
  \def\theequation{\Alph{section}.\arabic{equation}}}

\def\SymBoxes#1#2#3#4{\newdimen\un@t \un@t#3%
\raisebox{#1}{\rule{#2\un@t}{#4}\hskip-#2\un@t
\@tempdimb\un@t \advance\@tempdimb by-#4\@tempcntb#2\relax%
\@whilenum{\@tempcntb>0}\do{
\rule{#4}{\un@t}\hskip\@tempdimb \advance\@tempcntb by\m@ne}%
\hskip-#2\un@t \rule[\un@t]{#2\un@t}{#4}%
\rule[\un@t]{#4}{#4}\hskip-#4
\rule{#4}{\un@t}}\hskip-#4}                

\newcommand{\beq}{\begin{equation}}
\newcommand{\eeq}{\end{equation}}

\begin{document}
\def\thefootnote{\fnsymbol{footnote}}


\begin{center}
{\large {\bf Getting the Most Neutrinos out of IsoDAR}}

\bigskip

\bigskip

{\large \noindent Emilio Ciuffoli${}^{1}$\footnote{ciuffoli@impcas.ac.cn}, Hosam Mohammed${}^{1,2}$\footnote{hosam@impcas.ac.cn}, Jarah
Evslin${}^{1,2}$\footnote{jarah@impcas.ac.cn}, Fengyi Zhao${}^{1}$\footnote{fengchu@impcas.ac.cn}  and Maksym Deliyergiyev${}^{1}$\footnote{deliyergiyev@impcas.ac.cn}}


\renewcommand{\thefootnote}{\arabic{footnote}}

\vskip.7cm

\vspace{0em} {\em  1) Institute of Modern Physics, CAS, NanChangLu 509, Lanzhou 730000, China\\
2) University of the Chinese Academy of Sciences, YuQuanLu 19A, Beijing 100049, China}

\vskip .4cm

\end{center}

\noindent
\begin{center} {\bf Abstract} \end{center}

\noindent
Several experimental collaborations worldwide intend to test sterile neutrino models by measuring the disappearance of antineutrinos produced via isotope decay at rest (IsoDAR).  The most advanced of these proposals have very similar setups, in which a proton beam strikes a target yielding neutrons which are absorbed by a high isotopic purity ${}^7$Li converter, yielding ${}^8$Li whose resulting decay yields the antineutrinos.  In this note, we use FLUKA and GEANT4 simulations to investigate three proposed modifications of this standard proposal.  In the first, the ${}^7$Li is replaced with ${}^7$Li compounds including a deuterium moderator.  In the second, a gap is placed between the target and the converter to reduce the neutron bounce-back.  Finally, we consider cooling the converter with liquid nitrogen.  We find that these modifications can increase the antineutrino yield by as much as 50 percent.  The first also substantially reduces the quantity of high purity ${}^7$Li which is needed.

\pagebreak

\section{Introduction}
\subsection{Motivation}
Various anomalies can be explained if one invokes sterile neutrinos.  Among the most intriguing of these is the LSND anomaly \cite{lsnd97,lsnd01}, in which $\overline{\nu}_e$ appeared in a detector located 30 meters away from a $\mu^+$ decay at rest $\overline{\nu}_\mu$ source.  While the appearance signal is in nearly $4\sigma$ of tension with a three flavor mixing model, it is easily explained in the presence of a single flavor of sterile neutrino.  However other appearance experiments in the same channel have produced negative \cite{karmen,icarus} or inconclusive \cite{miniboone} results.  Anomalous disappearance of $\overline{\nu}_e$ produced by a radioactive source in a detector \cite{gallex,sage}, called the Gallium anomaly, can also be explained with a single flavor of sterile neutrino with a mass of at least 1 eV.

The deficit of measured reactor antineutrinos with respect to state-of-the-art predictions~\cite{mueller,huber}, called the reactor anomaly \cite{mention}, can also be explained with a single sterile neutrino with a mass which may be as small as $0.1$ eV.  However in this case the shape of the observed spectrum \cite{renobump,renobump2,doublebump,dayabump} is in disagreement with the theoretical predictions.  It is more difficult for sterile neutrinos to explain this spectral deformation, and so many authors have instead argued that it results from a fault in the calculations of the theoretical spectra \cite{francesiteor,dayateor}.  Consistent with this interpretation of the deficit is recent evidence from Daya Bay that the contributions of the primary fission isotopes to the deficit are not proportional to their abundances~\cite{dayaev}.


Such massive sterile neutrinos are in some tension with Planck CMB data \cite{planck2015}.  However, assuming the standard $\Lambda$CDM cosmological model, the Planck data is in 3$\sigma$ of tension with an every growing list of measurements, from Lyman $\alpha$ forest baryon acoustic oscillations \cite{lyalpha} to the Hubble constant as determined by the local distance ladder \cite{riesslocal}.  While massive sterile neutrinos alone cannot eliminate these tensions \cite{grant1,grant2}, they do ease the tension by increasing the uncertainties reported by Planck \cite{planck2015} and they may be part of a larger solution involving dynamical dark energy \cite{hong,gongbo17,silk17}.

\subsection{Sterile Neutrino Searches}
Often motivated by these anomalies, there have been a number of proposed experiments which will search for sterile neutrinos.  Most of these proposals search for sterile neutrinos in the disappearance channel $\overline{\nu}_e\rightarrow\overline{\nu}_e$.  Experiments using reactor neutrinos are at an advanced stage \cite{neut16sterile}: some have completed runs \cite{neos}, are already running \cite{nucifer,stereo,danss}, have a running prototype detector module \cite{solid}, or are under construction \cite{prospect}.

As explained in Ref.~\cite{giuntireview}, $\beta$ decay experiments are sensitive to sterile neutrino oscillations if the sterile neutrino is sufficiently massive.  In Ref.~\cite{giunti17}, the authors used the $\beta$ decay experiments Mainz \cite{mainz} and Troitsk \cite{troitsk}, fit together with $\nu_e$ and ${\overline{\nu}}_e$ disappearance data, to derive a 2$\sigma$ upper bound on the sterile neutrino mass squared splitting of 29 eV${}^2$.  

CPT invariance implies that the $\nu_e$ and $\overline{\nu}_e$ survival probabilities are equal, as are those of $\nu_\mu$ and $\overline{\nu}_\mu$.  If furthermore there is only a single flavor of sterile neutrino, then the $\nu_e$ and $\nu_\mu$ survival probabilities as well as the $\nu_e\leftrightarrow\nu_\mu$ transition rates are determined by only two parameters, $U_{e4}$ and $U_{\mu 4}$.  This implies that anomalies in the $\overline{\nu}_e$ appearance and disappearance channels, when combined, place constraints on $\nu_\mu$ disappearance and vice versa.  So far there is no evidence for $\nu_\mu$ disappearance at shorter baselines than would be expected from the standard mixing of the 3 active neutrinos.  This lack of evidence therefore constrains $\overline{\nu}_e$ appearance and disappearance due to a single flavor of sterile neutrino.  In particular, Ref.~\cite{giunti17} has shown that recent limits on short baseline $\nu_\mu$ disappearance by MINOS \cite{minos16} and IceCube \cite{icecube16}, when incorporated into a global fit, exclude sterile neutrino mass squared splittings below 1 eV${}^2$ at the $2\sigma$ level.

According to Ref.~\cite{giunti17}, in a large part of the remaining parameter space, {{disappearance channel sterile neutrino searches using Isotope Decay At Rest (IsoDAR) are particularly sensitive.}}  To our knowledge in all such proposals  a neutron source produces neutrons which are absorbed by ${}^7$Li.  This produces ${}^8$Li whose decay produces $\overline{\nu}_e$ with a well known energy spectrum, extending to 13 MeV, with an average energy of 6.5 MeV.   The neutron source is usually a high intensity accelerator \cite{daed,noiads,russijuno}, but can also be an intense neutron emitting isotope \cite{isodarcorea} or a nuclear reactor \cite{russireattore}.  This canonical setup was first proposed in Ref.~\cite{mikaelian}, in which the neutron source was a ``special nuclear reactor".  After half a century, this idea has come full circle with proposals to use an accelerator driven system (ADS) subcritical reactor as the neutron source.

The most advanced proposals have been made by the DAE$\delta$ALUS collaboration.  The neutron sources in these proposals are cyclotrons which are under development as part of an ADS reactor and active interrogation program \cite{daed}.  The cyclotrons accelerate a high intensity proton beam or a H${}_2^+$ beam which is then dissociated into a proton beam.  The proton beam energy is 60 MeV with a current of 10 mA.  The protons strike a beryllium target, creating spallation neutrons.  In some cases the neutrons exit from the target into a heavy water moderator \cite{daed12,daedjuno}, while in some cases there is no moderator \cite{daedkam}.  In either case, they then enter into a sleeve containing isotopically pure ${}^7$Li, sometimes in the compound FLiBe (Li${}_2$BeF${}_4$) \cite{daedkam}, where they are absorbed yielding ${}^8$Li.  The eventual ${}^8$Li decay creates $\overline{\nu}_e$ with a well-known energy spectrum \cite{pr50,livelli,2015li8}.

IsoDAR disappearance channel experiments have several advantages over reactor neutrino experiments.  First, the fact that the spectrum is fairly well known reduces a major source of error in reactor neutrino experiments.  Second, after weighting by the inverse $\beta$ decay cross section, 87\% of the $\overline{\nu}_e$ have energies above 6 MeV.  At these relatively high energies the accidental background that plagues reactor experiments is reduced by two orders of magnitude~\cite{nucifer}.  Third, this higher energy means that the same distance to energy ratio $L/E$, and so the same oscillation phase, is achieved at greater distances.  As a result more shielding can be added and, more crucially, distance resolution requirements are weakened.  This also allows access to higher sterile neutrino masses.

To save money, all IsoDAR proposals use existing infrastructure, either the accelerator or the detector.  In particular DAE$\delta$ALUS has provided a detailed proposal for an IsoDAR experiment at KamLAND \cite{daedkam} and the JUNO collaboration has also included such an experiment in its plan \cite{daedjuno}. In addition in Refs.~\cite{noiads,wenlong} IsoDAR experiments have been proposed using the LINACs that are being built for China's Accelerator Driven System (ADS) subcritical reactor project.  In particular, a 25 MeV, 10 mA proton accelerator will be completed this year, although for now perhaps only at 5 mA, and a 250-600 MeV, 10 mA accelerator called China Initial ADS (CI-ADS) will be completed in 2022, with civil engineering beginning this year.  

\subsection{Summary of results}
The target stations of all four of the above IsoDAR proposals are quite similar. In this paper we will present the results of our simulations of various modifications of these target stations.  Our objective is to determine to what extent various modifications affect the $\overline{\nu}_e$ yield.  In particular we will not be interested here in how these modifications may be implemented, in the effect on the sensitivity to sterile neutrino searches or other science goals or even on the absolute normalization of the $\overline{\nu}_e$ flux.  As a result our study is quite straightforward, lending itself to simulation with FLUKA and also with GEANT4 using the physics list FTFP$\_$BERT$\_$HP below 250 MeV and QGSP$\_$BIC$\_$HP at 250 MeV.

These results will lead us to three main conclusions:
\newcounter{outline}
\begin{list}{\arabic{outline})} {\usecounter{outline} \setlength{\leftmargin}{0cm}\setlength{\itemsep}{.0cm}}
\item Mixing the moderator and the ${}^7$Li converter \cite{russi90} increases the $\overline{\nu}_e$ yield by as much as 50\%.
\item In this mixed case, for a sufficiently large converter, the $\overline{\nu}_e$ rate can be estimated quite precisely using only the overall normalization of the neutron yield, the cross sections and a simple analytic formula.
\item In the case of the 250 MeV CI-ADS beam with the preferred $W$ target, a gap between the target and the converter can reduce the neutrons lost by bounce-back into the target by as much as 30\%.
\end{list}

We will organize our discussion by dividing the simulation into parts.  First, in Sec.~\ref{transez} we will describe the transport of neutrons in the converter and the resulting isotope production.  This will be done in an idealized setting with no proton beam or target and a monochromatic neutron source.  In Sec.~\ref{prodsez} we will consider the proton beam striking the target and the production of spallation neutrons.  Finally in Sec.~\ref{pienosez} we will describe our full simulations, from the proton beam to ${}^8$Li production.

\section{Neutron Transport} \label{transez}
In this section we will consider the simpler problem of calculating the $\overline{\nu}_e$ rate given a monochromatic neutron source in the center of the converter which is surrounded by a graphite reflector.  We will see that at the relevant energies, the $\overline{\nu}_e$ yield has little dependence on the neutron energy and, for a large enough converter, can easily be estimated analytically.  We feel that the results in this section provide an intuitive understanding of the results of the full target station simulations which will be presented in Sec.~\ref{pienosez}.

\subsection{Analytic Results}

In this subsection we will make the crude approximation that the converter is infinite in extent and try to anticipate the expected behavior of the neutrons in a simplified random walk model.  All of the converters which we will consider consist of H, D, Be, ${}^6$Li, ${}^7$Li, O and F, with O having the isotopic abundances found in nature.  At the energies of interest the absorption cross sections are inversely proportional to the neutron velocity, and the elastic scattering cross sections are energy-independent up to about 1 MeV, where they begin to fall.  Above 100 keV there are also some resonances which affect these cross sections considerably.  These resonances will not be included in our analytical model, although of course they are incorporated into the libraries used by our simulations.

The vast majority of our neutrons will have initial energies between 100 keV and 5 MeV, and so we approximate the elastic scattering cross sections to be energy-independent.   We will use the cross sections from Ref.~\cite{neutronnews} which are summarized in Table~\ref{crosstab}.  All absorption cross sections are reported at $0.025$\ eV, to convert to other energies it suffices to scale inversely by the mean neutron velocity.

Let $\rho_i$ be the number of isotopes of type $i$ per unit volume and let $\sigma^i_{\rm{elastic}}$ be the elastic scattering cross section for a neutron on an isotope of type $i$.  Considering only elastic scattering, as is reasonable before the neutrons are thermalized, the mean free path is
\beq
\lambda=\frac{1}{\sum_i \rho_i \sigma^i}.
\eeq

We will consider a simple model of neutron moderation in which the neutron loses 40\% of its energy during each collision with a D, and no energy during the other collisions.  In particular we will ignore elastic scattering with H which will be quite rare in the cases that follow as a result of the high isotopic purity of D.  The probability that a given elastic scattering involves a D is equal to
\beq
p=\rho_D\sigma^D \lambda.
\eeq
Therefore roughly every $1/p$ elastic scatterings there is an elastic scattering with a D.  Let us assume that at each of these $1/p$ scatterings, as the target is much heavier than the neutron, the neutron's direction is randomized and so the neutron follows a 3-dimensional random walk.  Therefore the neutron will travel, on average, a distance
\beq
d_{D}=\sqrt\frac{2}{3\pi p}\lambda=\sqrt{\frac{2\lambda}{3\pi\rho_D\sigma^D}}
\eeq
in any given direction between two elastic scatterings with D.

A neutron which is created with an energy of $E$ MeV will thermalize to room temperature after -ln($4E\times 10^7$)/ln(0.6) collisions with D.  Now we will make the poor approximation that the neutron randomizes its direction when scattering with D, effectively ignoring the recoil of the deuteron. Then during these collisions, it will travel an expected distance of
\beq
d_{\rm{therm}}=\sqrt{\frac{2\lambda {\rm{ln}}(4E\times 10^7)}{{3\pi{\rm{ln}}(5/3)}\rho_D\sigma^D}}
\eeq
in a given direction.  For example, one expects neutrons to travel a distance $d_{\rm{therm}}$ in the radial direction of a wide, hollow, cylindrical converter before thermalization.  We have numerically simulated random walks with and without D recoil, assuming that the converter consists entirely of deuterons and have found that including the D recoil increases $d_{\rm{therm}}$ by 40\%.  A smaller correction can be expected in compounds which include heavier isotopes.

\begin{table}[t]
\centering
\begin{tabular}{|c|l|l|l|l|l|l|l|}
&H&D&${}^6$Li&${}^7$Li&Be&O&F\\
\hline\hline
$\sigma_{\rm{elastic}}$ (barns)&82.03&7.64&0.97&1.4&$7.63$&4.232&4.018\\
\hline
$\sigma_{\rm{abs}}$ (barns) &0.3326&$5.19\times 10^{-4}$&940&0.0454&$0.0076$&$1.9\times 10^{-4}$&$0.0096$\\
\hline
\end{tabular}   
\caption{Elastic scattering and absorption cross sections of various isotopes. \label{crosstab}}
\end{table}

Finally we make the reasonable approximation that neutrons can only be absorbed after they have thermalized.  The probability that an interaction after thermalization leads to absorption is
\beq
p_{\rm{abs}}=\frac{\sum_i\rho_i\sigma^i_{\rm{abs}}}{\sum_i\rho_i\left(\sigma^i_{\rm{elastic}}+\sigma^i_{\rm{abs}}\right)}. \label{pabs}
\eeq
Therefore one expects that after thermalization a neutron will scatter $1/{p_{\rm{abs}}}$ times before being absorbed, during which it travels an expected distance of
\beq
d_{\rm abs}=\sqrt\frac{2}{3\pi p_{\rm abs}} \lambda\label{dabs}
\eeq
in a given direction, for example in the radial direction.  We will make the rough approximation that $\lambda$ is the same before and after thermalization.

Thermalized neutrons in general will be absorbed in the converter.  The probability that a thermalized neutron is absorbed by ${}^7$Li and therefore yields a $\overline{\nu}_e$ is
\beq
p_\nu=\frac{\left(\rho_{{}^7\rm Li}\right)\left(\sigma^{{}^7\rm Li}_{\rm abs}\right)}{\sum_i\rho_i\sigma^i_{\rm{abs}}}. \label{pn}
\eeq

\begin{table}[t]
\centering
\begin{tabular}{|c|l|l|l|l|l|}
&Li&LiOD&LiOD$\cdot$D${}_2$O&solution&FLiBe\\
\hline\hline
density (gm/cm${}^3$)&0.534&1.52 \cite{cdc}&1.62 \cite{cdc}&1.1&1.94 \cite{arwif}\\
\hline
$\rho_{\rm H}$ (cm${}^{-3}$) &0&$3.66\times 10^{20}$&$6.50\times 10^{20}$&$6.16\times 10^{20}$&$0$\\
\hline
$\rho_{\rm D}$ (cm${}^{-3}$) &0&$3.62\times 10^{22}$&$6.44\times 10^{22}$&$6.10\times 10^{22}$&$0$\\
\hline
$\rho_{{}^6\rm Li}$ (cm${}^{-3}$) &$4.59\times 10^{18}$&$3.66\times 10^{18}$&$2.17\times 10^{18}$&$3.24\times 10^{17}$&$2.36\times 10^{18}$\\
\hline
$\rho_{{}^7\rm Li}$ (cm${}^{-3}$) &$4.59\times 10^{22}$&$3.66\times 10^{22}$&$2.17\times 10^{22}$&$3.24\times 10^{21}$&$2.36\times 10^{22}$\\
\hline
$\rho_{\rm Be}$ (cm${}^{-3}$) &$0$&$0$&$0$&$0$&$1.18\times 10^{22}$\\
\hline
$\rho_{\rm O}$ (cm${}^{-3}$) &$0$&$3.66\times 10^{22}$&$4.34\times 10^{22}$&$3.58\times 10^{22}$&$0$\\
\hline
$\rho_{\rm F}$ (cm${}^{-3}$) &$0$&$0$&$0$&$0$&$4.72\times 10^{22}$\\
\hline
\end{tabular} 
\caption{Densities and isotope number densities in various converters.  The densities have been rescaled from the original references to reflect the desired isotope compositions. \label{denstab}}
\end{table}

We will be interested in five different converter materials.  In each the D will be 99\% isotopically pure (99\% mole fraction), with the remaining 1\% being H.  Also the ${}^7$Li will be 99.99\% isotopically pure, with the remaining 0.01\% consisting of ${}^6$Li.  In practice the vendors with whom we have spoken offer much lower prices if there are some other impurities, however these other impurities are irrelevant here due to their low neutron absorption cross sections.  The five materials are pure metallic Li, LiOD, LiOD$\cdot$D${}_2$O, a heavy water solution which is 11.6\% LiOD by mass and finally FLiBe.  The first material has been chosen in most IsoDAR proposals \cite{daed12}, the last in IsoDAR at KamLAND \cite{daedkam}, and the others have been suggested in Ref.~\cite{russi90}.

\begin{table}[t]
\centering
\begin{tabular}{|c|l|l|l|l|l|}
&Li&LiOD&LiOD$\cdot$D${}_2$O&solution&FLiBe\\
\hline\hline
$\lambda$ (cm)&$15.5$&$1.95$&$1.32$&$1.52$&$3.20$\\
\hline
$p$&$0$&$0.540$&$0.648$&$0.709$&$0$\\
\hline
$d_{\rm D}$ (cm)&$\infty$&$1.22$&$0.754$&$0.832$&$\infty$\\
\hline
$d_{\rm therm}$ (cm) $E=1$&$\infty$&$7.15$&$4.41$&$4.87$&$\infty$\\
\hline
$d_{\rm abs}$ (cm)&$23.8$&$8.92$&$9.25$&$21.9$&$13.4$\\
\hline
$p_\nu$&$0.326$&$0.317$&$0.300$&$0.208$&$0.280$\\
\hline
\end{tabular} 
\caption{Neutron transport properties of each converter \label{quantab}}
\end{table}

Each compound can have various densities and bulk densities depending on the crystalline structure and/or preparation.  We have chosen not to optimize these densities, but rather we have chosen the densities which appear most often on the web pages of vendors, as these are likely to be the most readily available.   These densities were then rescaled to the isotope specifications of interest for our study.  The results are summarized in Table~\ref{denstab}.  In the case of the heavy water solution we simply used the density of heavy water.  The density of metallic lithium has an appreciable temperature dependence, and we have used a density corresponding to room temperature.

Combining the number densities $\rho_i$ in Table~\ref{denstab}, together with the cross sections in Table~\ref{crosstab}, one can now evaluate the various quantities above for each converter.  The results are shown in Table~\ref{quantab}.  As we have made the approximation that only D moderates, the thermalization distance for the metallic Li and the FLiBe converters can not be evaluated.  In the other cases, the longest distance is the absorption distance $d_{\rm abs}$ which is approximately 9 cm for LiOD and LiOD$\cdot$D${}_2$O and 22 cm for the solution.   

As can be seen in Eq.~(\ref{dabs}), the absorption distance is determined by two quantities: the mean free path $\lambda$ and the probability of absorption per collision $p_{\rm{abs}}$.  ${}^6$Li is the dominant absorber in each case, so a high concentration of Li leads to a high $p_{\rm{abs}}$ and so a low $d_{\rm{abs}}$.  For example, the solution has a low concentration of Li and a large $d_{\rm{abs}}$.  The exception to this rule is metallic Li, whose low $\sigma_{\rm{elastic}}$ leads to a large $\lambda$.  As a result, neutrons can travel long distances unimpeded in metallic Li, and so it has the longest $d_{\rm{abs}}$.  Similarly $d_{\rm{therm}}$ is in general lowest for the compounds with the highest concentrations of D, which is efficient both for slowing and for scattering neutrons.  However LiOD$\cdot$D${}_2$O thermalizes neutrons slightly more quickly than the solution due to its higher density.


One expects that the ${}^8$Li production, and so the $\overline{\nu}_e$ production per neutron will saturate to $p_\nu$ when the converter radius is sufficiently large.   The saturation value $p_\nu$ is much less for the solution, but comparable for the other converters. 

The fact that $p_\nu$ is close to $1/3$ is easy to understand.  Apart from the solution, whose high D content implies that 30\% of neutrons are absorbed by H, in all other materials Li is responsible for at least 85\% of neutron absorption.  There is $10^4$ times more ${}^7$Li than ${}^6$Li in each case, but $\sigma_{\rm abs}$ of ${}^6$Li is $2\times 10^4$ times higher than that of ${}^7$Li.  Therefore ${}^6$Li absorbs twice as many neutrons as ${}^7$Li, leaving about $1/3$ of the neutrons for ${}^7$Li.   Metallic Li has the highest value of $p_\nu$ as it contains no other neutron absorbers, whereas the solution has the lowest due to its high H content.  If on the other hand the ${}^7$Li purity is increased to 99.995\% as in Ref.~\cite{daedkam}, then the same argument implies that $p_\nu$ will be about $1/2$, corresponding to a 50\% increase in the $\overline{\nu}_e$ yield.  We hope that the optimizations described in this note may lead to a smaller, more efficient converter which in turn would allow, at the same price, a higher isotopic purity of Li and so a higher $\overline{\nu}_e$ yield.

\subsection{Simulation Results} \label{neutsimsez}

{ {The above analytic model uses crude approximations to provide a qualitative understanding of the thermalization and absorption.  We will now remove those approximations and report the quantitative results of our neutron transport simulations.}}

We have simulated neutron transport using all of these converters with FLUKA \cite{fluka} and some of these also with GEANT4 \cite{geant}.  Each FLUKA configuration was simulated with at least $10^5$ monochromatic neutrons per energy, meaning that statistical fluctuations are negligible.  For this study, our configurations consist of concentric cylinders.  In the center is a vacuum with a 10 cm radius and a length of 20 cm.  In the case of metallic lithium, following the DAE$\delta$ALUS proposal \cite{daed12}, this is surrounded by 5 cm of heavy water on each side.  Next is the converter, which extends 10$n$ cm beyond the vacuum where we have run simulations for integral values of $n$.  In the case of the solution instead we consider 40, 80, 100, 120, 140, 160 and 180 cm of extension beyond the vacuum.  In every case this is surrounded by 60 cm of graphite reflector on each side.  For simplicity we have not included cooling systems.

\begin{figure} 
\begin{center}
\includegraphics[width=3.2in,height=1.5in]{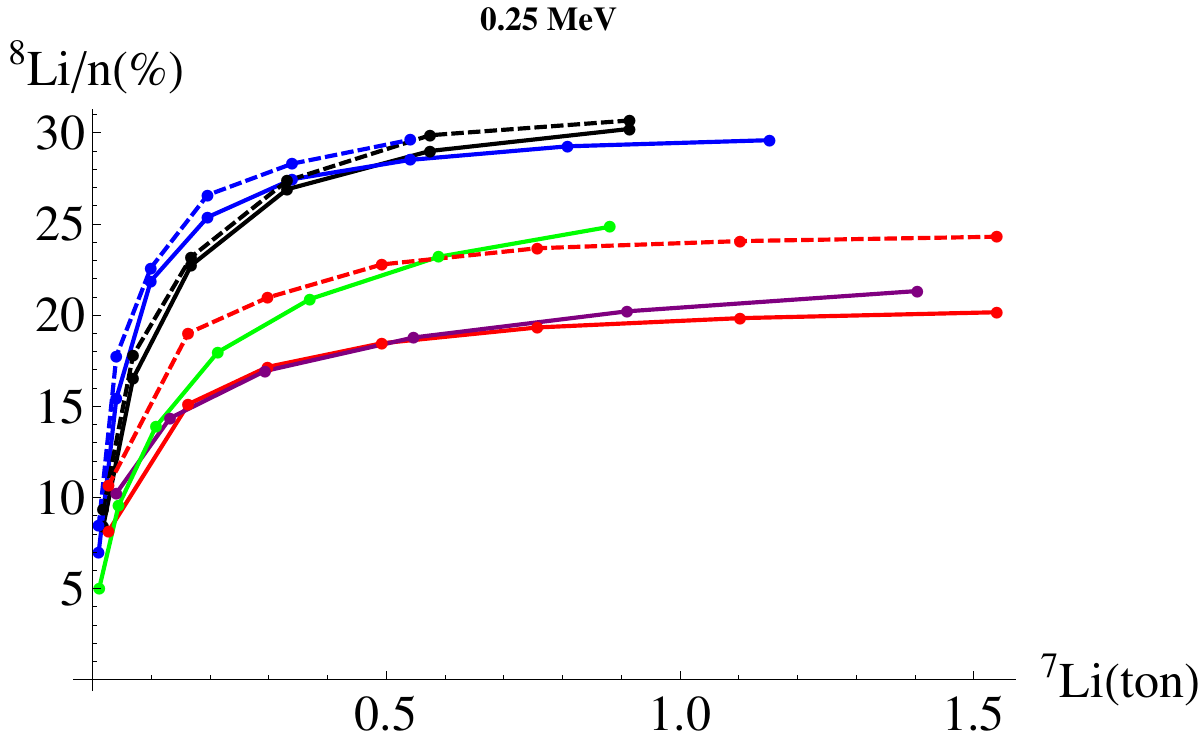}
\includegraphics[width=3.2in,height=1.5in]{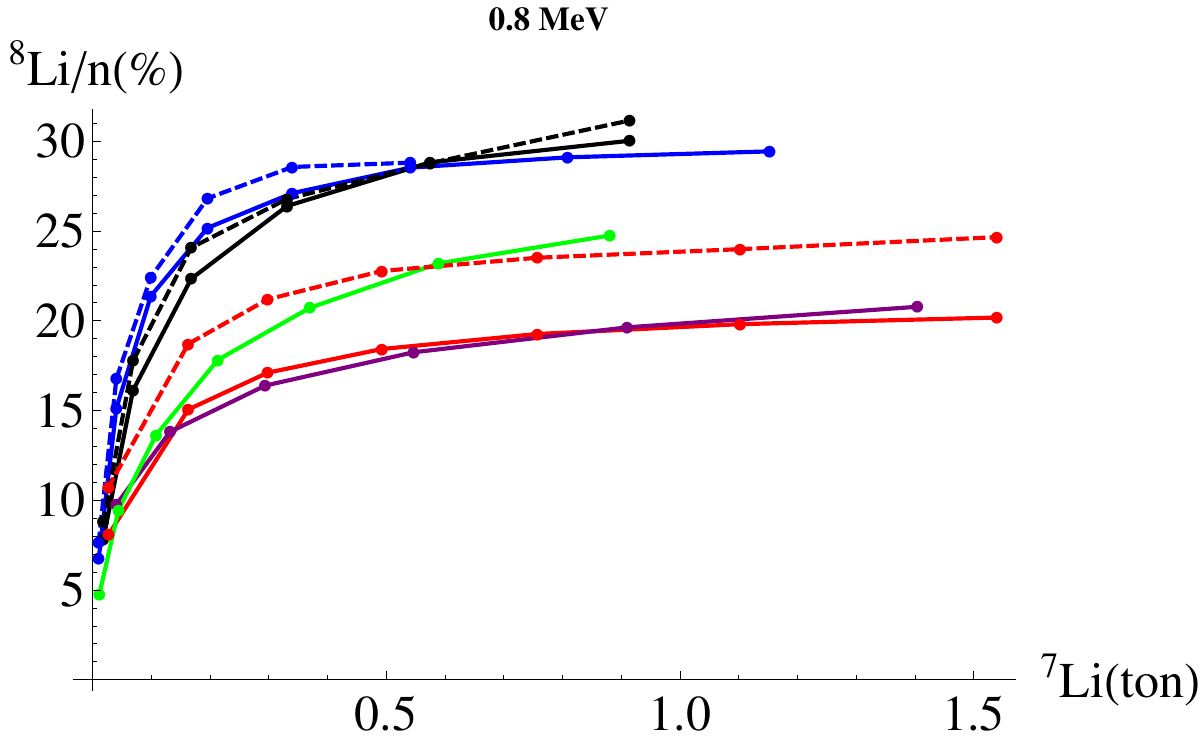}
\includegraphics[width=3.2in,height=1.5in]{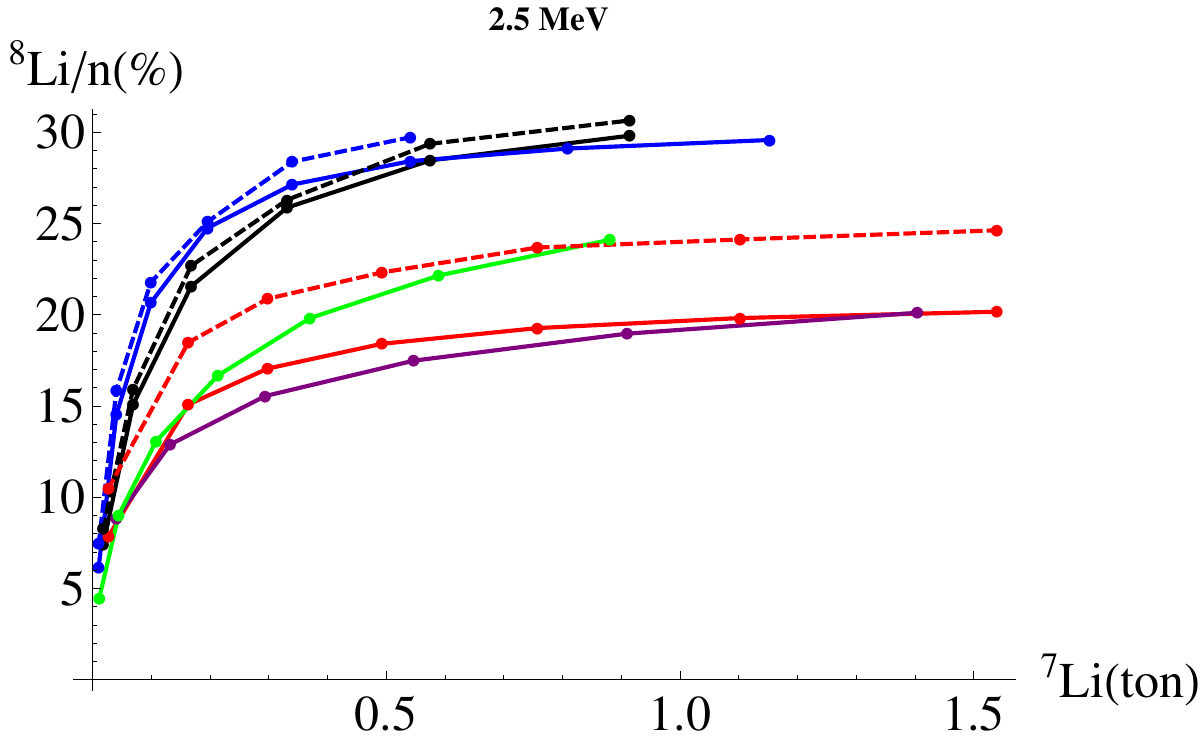}
\includegraphics[width=3.2in,height=1.5in]{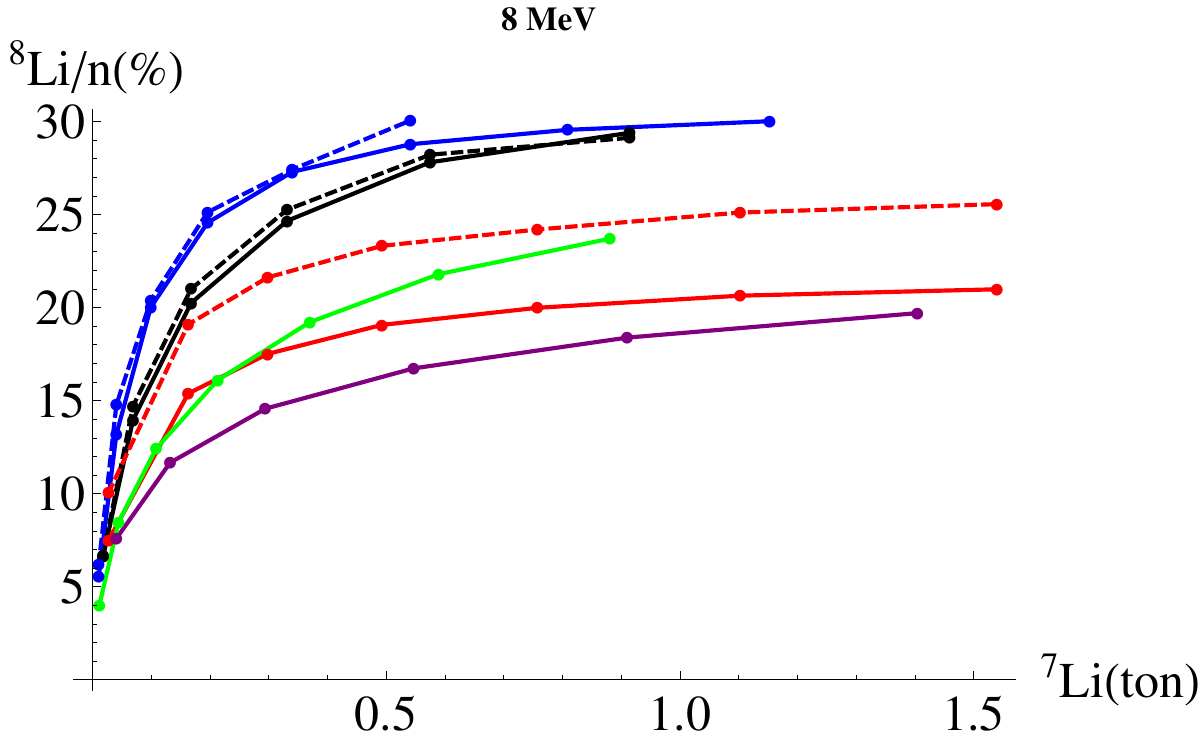}
\includegraphics[width=3.2in,height=1.5in]{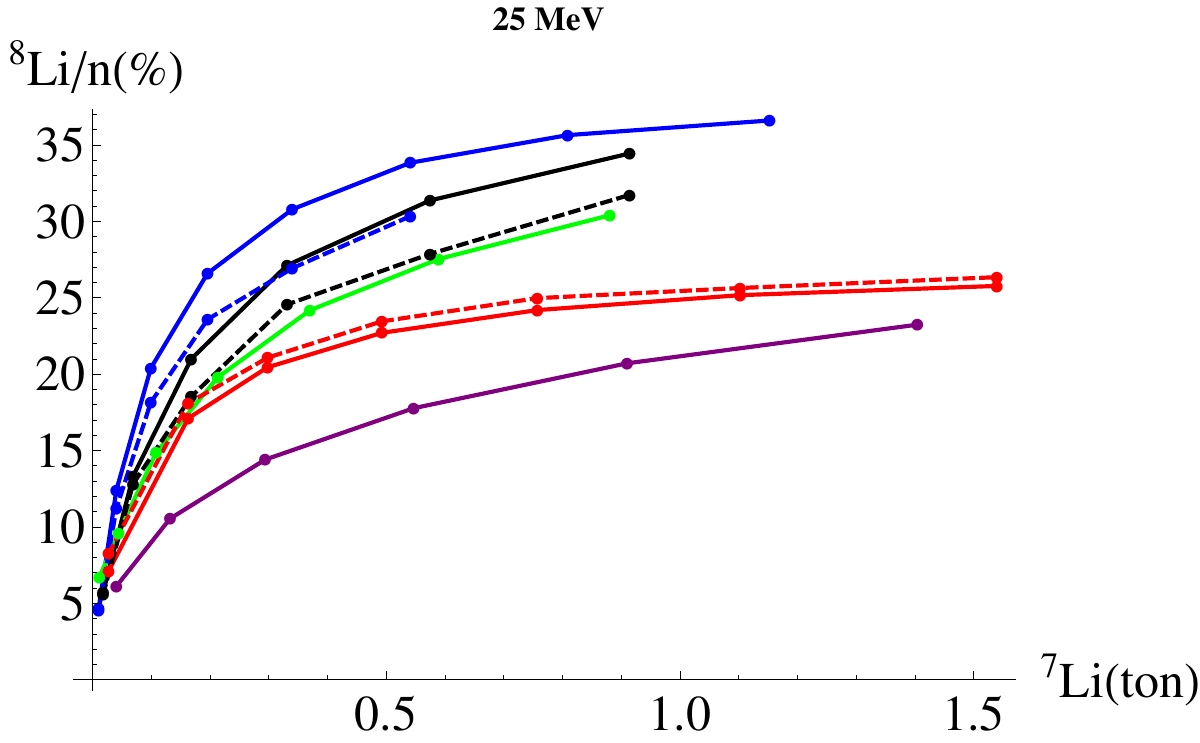}
\includegraphics[width=3.2in,height=1.5in]{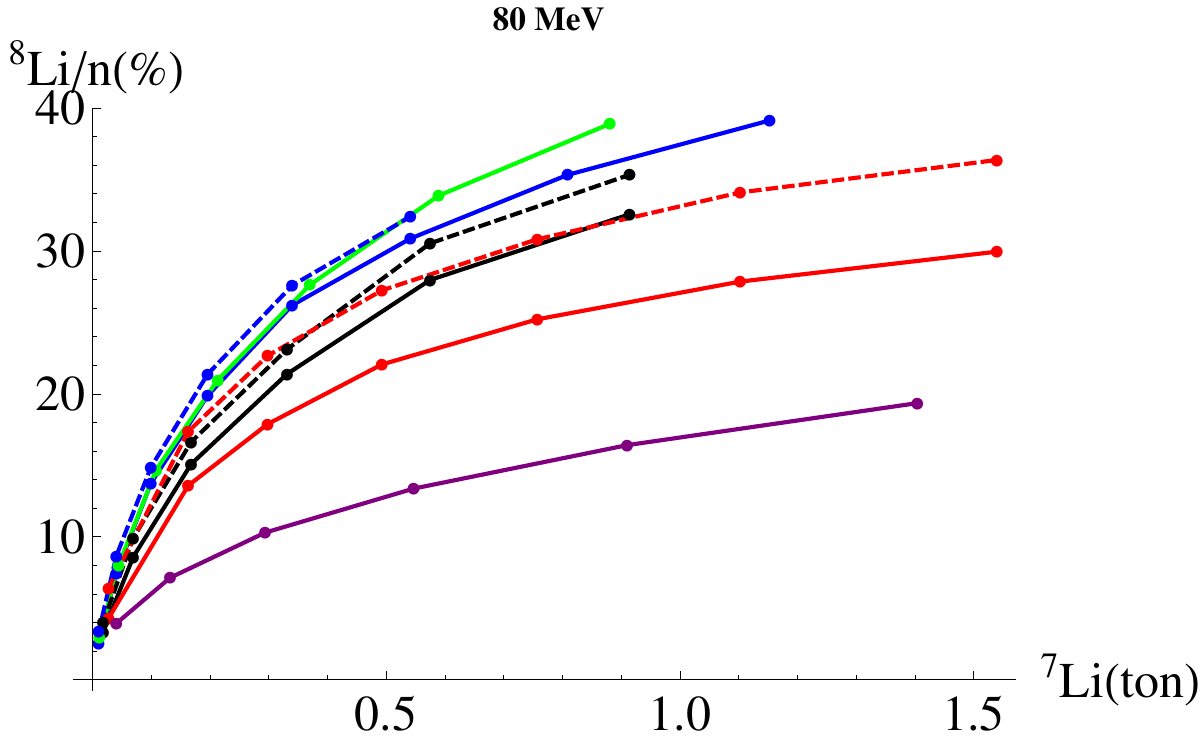}
\caption{The ${}^8$Li/neutron ratio for monochromatic neutrons at various energies.  The black, blue, red, purple and green curves represent the LiOD, LiOD$\cdot$D${}_2$O, solution, metallic Li and FLiBe converters respectively.  Solid curves were produced with FLUKA and dashed curves with GEANT4.  The horizontal axis is the mass of the ${}^7$Li.}
\label{trasfig}
\end{center}
\end{figure}

\begin{figure} 
\begin{center}
\includegraphics[width=3.2in,height=1.5in]{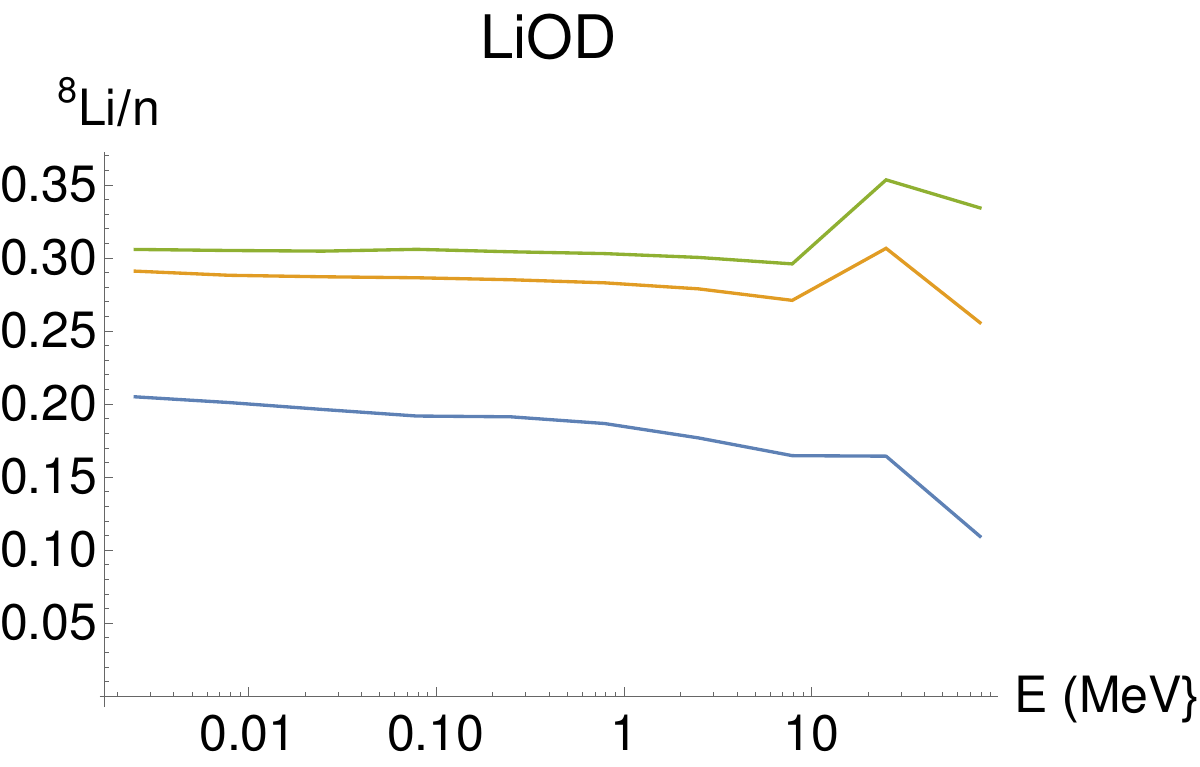}
\includegraphics[width=3.2in,height=1.5in]{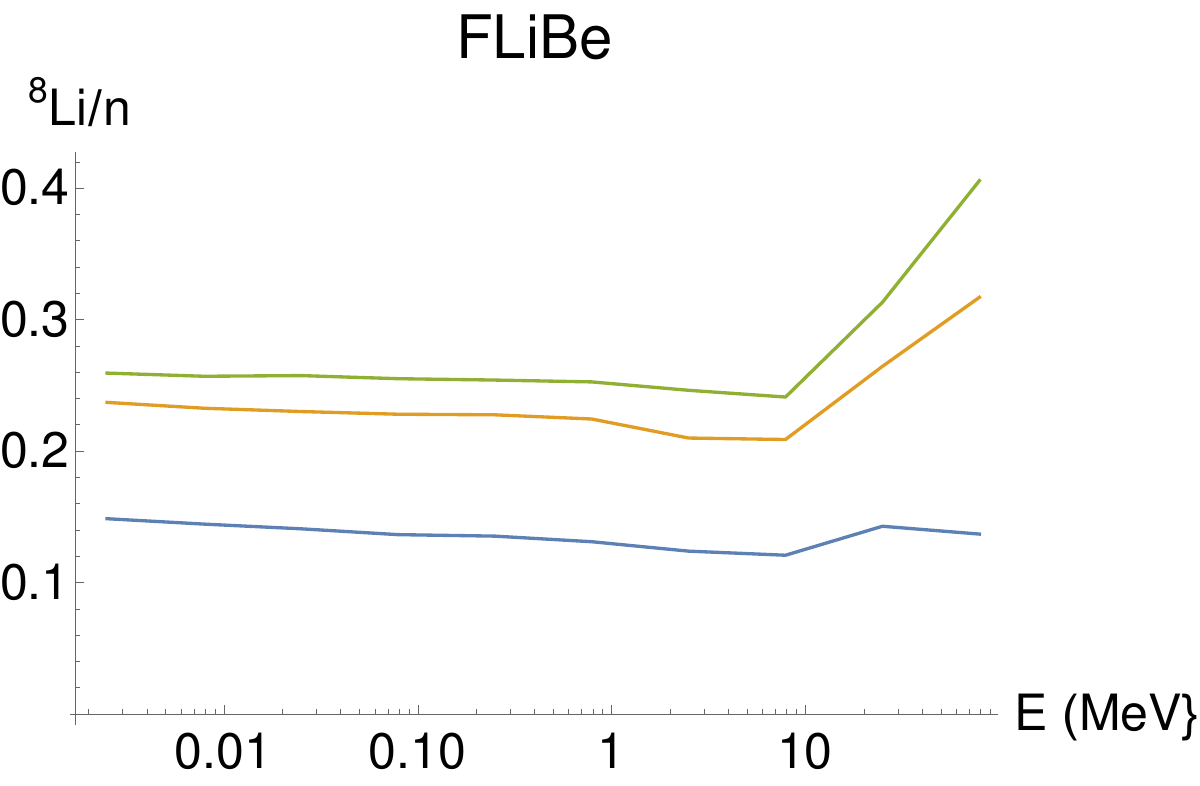}
\caption{The ${}^8$Li/neutron ratio for monochromatic neutrons at various energies in LiOD (left) and FLiBe (right).  The blue, yellow  and green curves represent the 0.1 tons, 0.5 tons and 1 ton of ${}^7$Li respectively.}
\label{mfissofig}
\end{center}
\end{figure}

The results of our simulations, for neutrons at energies of $0.25$ MeV, $0.8$ MeV, $2.5$ MeV, $8$ MeV, 25 MeV and 80 MeV, are shown in Fig.~\ref{trasfig} for various quantities of ${}^7$Li.   One may observe that, below 10 MeV, the ${}^8$Li production efficiency, or equivalently the $\overline{\nu}_e$ production efficiency, is essentially independent of the energy.  In the case of LiOD and FLiBe this is shown explicitly in Fig.~\ref{mfissofig}.  Above 10 MeV there are two competing effects.  First, the lower neutron elastic cross section reduces the ${}^8$Li production efficiency, as more neutrons escape.  This effect is largest for small converters, such as that represented by the blue curve.  In fact, the thermalization distance scales logarithmically with the energy and so in converters whose dimensions are of order the thermalization length, the ${}^8$Li yield is slightly energy dependent at all neutron energies.  Second, neutron multiplication increases the efficiency.  The latter effect is dominant in FLiBe as ${}^9$Be multiplies more efficiently than D due to the lower energy cost to remove a valence neutron from ${}^9$Be with respect to D.  Therefore in general FLiBe is better for very high energy neutrons.  

We will see below that for the beams considered here, the number of neutrons with energies above 10 MeV is negligible and so neutron multiplication will be inconsequential.  However, a high energy deuteron beam will create forward neutrons at half of the deuteron energy.  Thus an IsoDAR experiment at a deuteron beam may want to use a FLiBe converter in the forward direction from the target.  A 50 MeV, 10 mA deuteron beam is now being built at a user facility in Ningde, China and an upgrade to 200 MeV is foreseen.  A hybrid converter, consisting of FLiBe in the forward the direction and a D rich compound elsewhere, could provide an optimal design for an IsoDAR experiment at this beam.

\begin{figure} 
\begin{center}
\includegraphics[width=3.2in,height=1.5in]{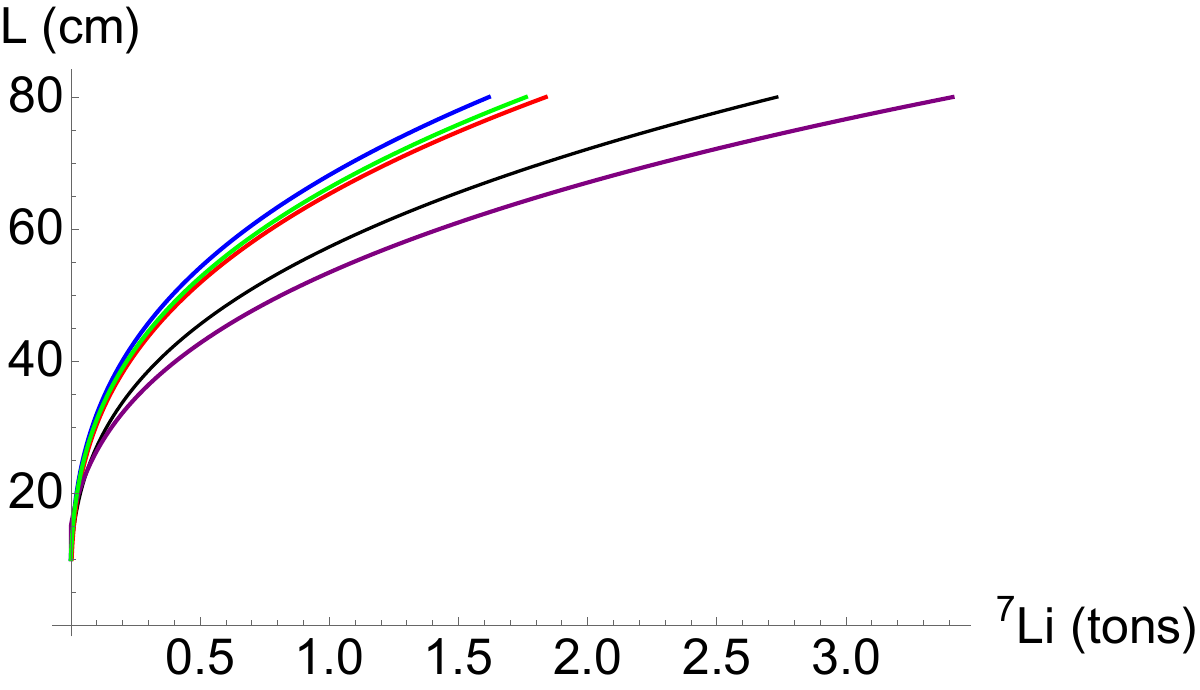}
\caption{The target station radius, including the detector and the converter as a function, of the ${}^7$Li mass.  The black, blue, red, purple and green curves represent the LiOD, LiOD$\cdot$D${}_2$O, solution, metallic Li and FLiBe converters respectively.  The radius is divided by two in the case of the solution, for clarity of the plot.}
\label{massalifig}
\end{center}
\end{figure}

The cost of the converter is driven by the pure ${}^7$Li and so it is reasonable to compare converters at fixed ${}^7$Li mass.  However, one can obtain the total mass from the ${}^7$Li mass by multiplying by 3.57, 6.43, 30.8, 1 or 7.07 for LiOD, LiOD$\cdot$D${}_2$O, solution, metallic Li and FLiBe converters respectively.   The radius of the target station as a function of the Li mass is shown in Fig.~\ref{massalifig}.  {{Such conversions may be of interest if mass and or space are more important constraints than costs, for example for underground configurations.}}

One may observe that on the right side of each panel in Fig.~\ref{trasfig}, as the mass is sufficient to thermalize and absorb the neutrons, in general the ${}^8$Li production per neutron, which is equal to the $\overline{\nu}_e$ production, reaches an asymptotic value.  This asymptotic value agrees well with $p_\nu$ calculated in Eq.~(\ref{pn}) in every case except for the GEANT4 simulation of the solution, which tends to be about 4\% too high.  Note that at energies above 2.5 MeV there is no asymptotic value, instead the ${}^8$Li production continues to increase as the converter size is increased.  This is because at these energies the neutrons have sufficient energy to break the D or Be in the converter, freeing more neutrons.  The neutron increase at these high energies is therefore a result of neutron multiplication.  While this neutron multiplication is significant at energies of 25 MeV and above, we will see that even 250 MeV protons create very few neutrons above 5 MeV, and so neutron multiplication in fact is insignificant in every case that we will consider.   In fact, we have run additional simulations with fictional converter materials that have a higher Be density and we have found that at 60 MeV they outperform all of the materials considered here.  

One exception to this argument is metallic Li, which does not arrive at an asymptotic value.  This is due to the fact that neutrons which are not already thermalized by the heavy water moderator require several meters of Li to thermalize, and so do not thermalize for the converter sizes that we have considered.  For this reason, proposals for IsoDAR experiments using metallic Li converters generally use several tons of Li.  Even with such larger converters, the asymptotic ${}^8$Li/n ratio will be less than $p_\nu$ due to absorption of neutrons in the moderator.

The main result of our paper is quite clear in every panel of Fig.~\ref{trasfig}.  The metallic Li converter outside of a heavy water moderator, which has been chosen at many IsoDAR experiments \cite{daed12}, in fact has an appreciably lower neutron yield than the two other solid converters considered when the Li mass is less than 1.5 tons.  This is true for every neutron energy, and so it will be true for every proton beam.  {{The effect is quite large and suggests that by mixing the moderator and the converter one may increase the $\overline{\nu}$ flux by as much as 50\%.  This result has been anticipated in Ref.~\cite{russi90}, although quantitatively our simulation results are quite different~\cite{russi15}.}}

\subsection{Thermalization and Absorption Distances}

To better understand the results of these simulations, in this subsection we will report the results of FLUKA simulations of a simplified geometry designed to determine the thermalization and absorption distances.  For this aim, we will consider solid, spherical converters of various radii with no reflector.  We will not consider metallic Li, as in IsoDAR proposals this is always used in conjunction with a moderator.  All neutrons will be created in the centre of the sphere at 1 MeV.

A neutron will thermalize or be absorbed inside of the moderator only if the maximal distance in its 3d random walk is less than the radius of the sphere.  The expected maximal distance in 3 dimensions exceeds the expected final distance in 1 dimension by a factor of $\sqrt{6}$.  Therefore one expects, for example, that most neutrons will thermalize if the radius exceeds $\sqrt{6}d_{\rm therm}$.   Similarly, one expects that most neutrons will be absorbed when the radius exceeds
\beq
r=\sqrt{6(d_{\rm therm}^2+d_{\rm abs}^2)}.
\eeq

In practice there are a number of corrections to this idealized estimate.  For example, the finite recoil, in particular of D in the target, will increase these distances by up to 40\%.   Also resonances in the neutron scattering cross section, which generally occur at 100s of keV and exceed the average cross section by an order of magnitude.  Our analytical calculation was performed with energy-averaged cross sections, however the resonances provide a considerable contribution to these average cross sections.  For example, they contribute nearly one third of the average elastic cross section of neutrons on ${}^7$Li.  As a result, neutrons lose energy quickly until about 100 keV, but most of the thermalization distance is traveled by neutrons below these resonances, where the elastic cross section is reduced.   This has the effect of increasing the true thermalization distance by several 10s of percent.  We have checked that these resonances are correctly implemented in both our FLUKA and GEANT4 simulations, in the former by considering scattering off of a thin target and in the {{latter}} by calculating the average trajectory length before a 1 MeV neutron reaches a specific energy.

\begin{figure} 
\begin{center}
\includegraphics[width=3.2in,height=1.5in]{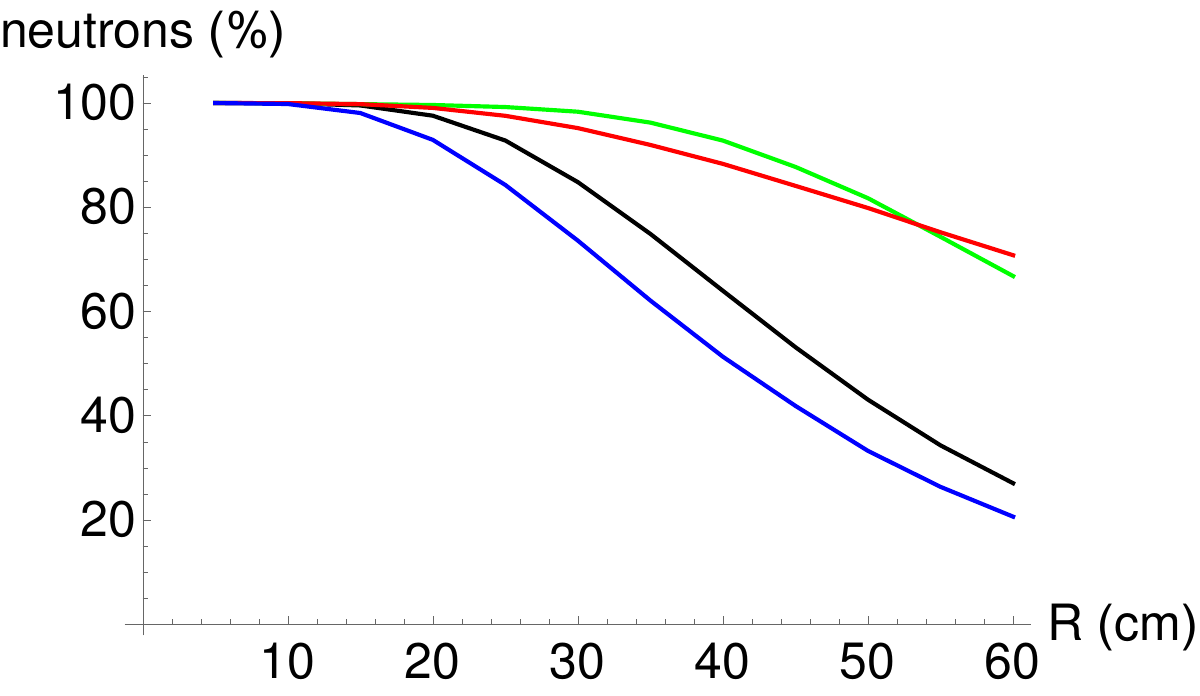}
\includegraphics[width=3.2in,height=1.5in]{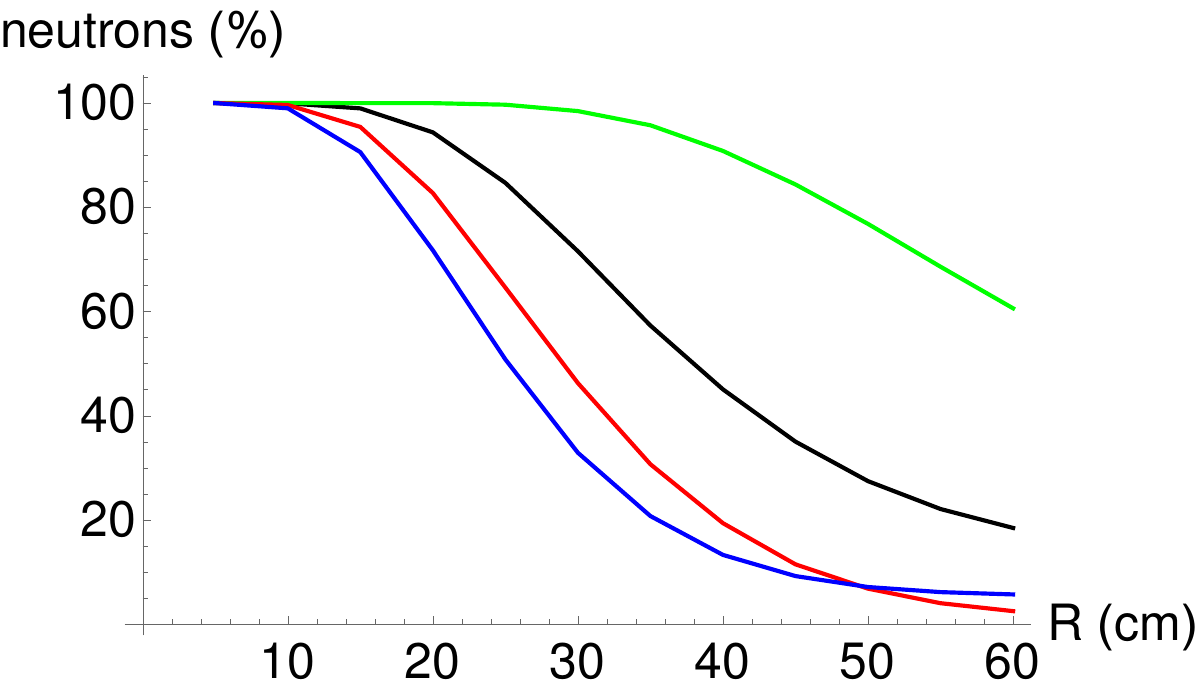}
\caption{(left) The percentage of initially monochromatic 1 MeV neutrons which escape a spherical converter.  (right) The percentage of initially monochromatic 1 MeV neutrons whose energy never falls below 0.028 eV before escaping or being absorbed.  The black, blue, red and green curves represent the LiOD, LiOD$\cdot$D${}_2$O, solution and FLiBe converters respectively.   The horizontal axis is the radius of the converter.}
\label{lunfig}
\end{center}
\end{figure}

Our results are summarized in Fig.~\ref{lunfig}.  On the left, we plot the fraction of neutrons which escape from a converter of various radii.  FLiBe and the solution are the worst absorbers, as the former is a poor moderator and so has a long thermalization distance while the later is a poor absorber, with $d_{\rm abs}$ equal to roughly 22 cm.  The best absorption is achieved by LiOD and LiOD$\cdot$D${}_2$O, which are adequate moderators and absorbers with $r$ equal to 28 and 25 cm respectively.   One can see that half of the neutrons are absorbed when the radius is about 40 cm.  This is somewhat larger than the value of $r$ found in the naive analytic model.

In the right panel we study the thermalization distance by plotting the percentage of neutrons which never reach $0.028$ eV before being absorbed or escaping the converter.  This is roughly equal to the percentage of neutrons which is not thermalized.  FLiBe is by far the worst performer, as it is a poor moderator.  On the other hand, most neutrons in LiOD$\cdot$D${}_2$O and the solution thermalize by 30 cm.  This is about twice the thermalization radius $\sqrt{6}d_{\rm therm}$ predicted in our naive analytic model, in part due to the finite nuclear recoils.  Therefore we see that while the analytic model successful predicts the relative performances of the converters, it somewhat underestimates the distances.

In the case of LiOD, the right panel yields a thermalization distance of 37 cm while the left panel yields a thermalization plus absorption (sum in quadrature) distance of 45 cm.  Thus the thermalization distance is greater than the absorption distance, in contrast with the analytic results which do not include the nuclear recoil contribution to the distances.    LiOD$\cdot$D${}_2$O is a better moderator and so these distances are 25 cm and 39 cm.  In this case the absorption distance exceeds the thermalization distance.  In both cases, the absorption distance yields a nontrivial contribution to the sum in quadrature.

\section{Neutron Production} \label{prodsez}

We will be interested in IsoDAR experiments which begin with a proton beam that strikes a target creating neutrons which are then absorbed by various isotopes.  Those absorbed by ${}^7$Li provide ${}^8$Li and so our $\overline{\nu}_e$ signal.  In Sec.~\ref{transez} we studied the second step of this experiment, the neutron transport and absorption.  In this section we will instead describe the first step, the production of neutrons at the target.

In each case, throughout this paper, the target will be 20 cm long with a 10 cm radius.  We have optimized the target dimensions in each case so as to maximize ${}^8$Li/p.  We have found that this standard size yields a ${}^8$Li production rate which is near the optimum value for the three proton beam energies considered, and so for simplicity we report only simulations with this fixed size.  The 25 MeV and 60 MeV proton beams always strike a Be target, whereas we consider heavy metal targets for the 250 MeV beam, as these produce a higher neutron yield above about 50 MeV.

We have simulated this production with GEANT4 and FLUKA and we have compared our results with experimental data at various energies up to 100 MeV \cite{tilquin,osipenko} and also with the simulations of Ref.~\cite{sim250} at 250 MeV.  In general we have found that the GEANT4 simulations yield 10-20\% less neutrons than experimental data and the FLUKA simulations 20-40\% less, whereas we found better than 1\% agreement with the simulations of Ref.~\cite{sim250}.  The deficit in neutron production in FLUKA arises entirely at low energies.  

\begin{figure} 
\begin{center}
\includegraphics[width=3.2in,height=1.5in]{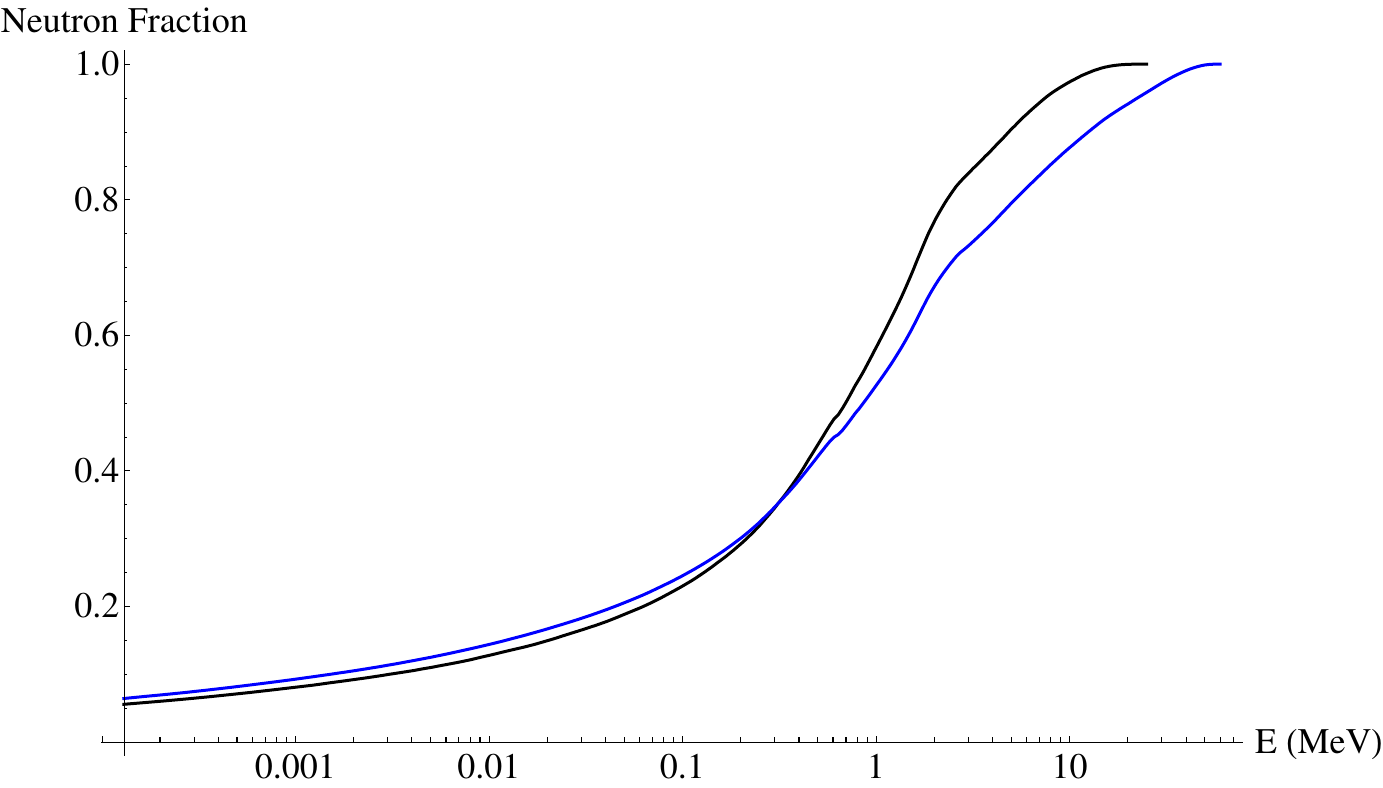}
\includegraphics[width=3.2in,height=1.5in]{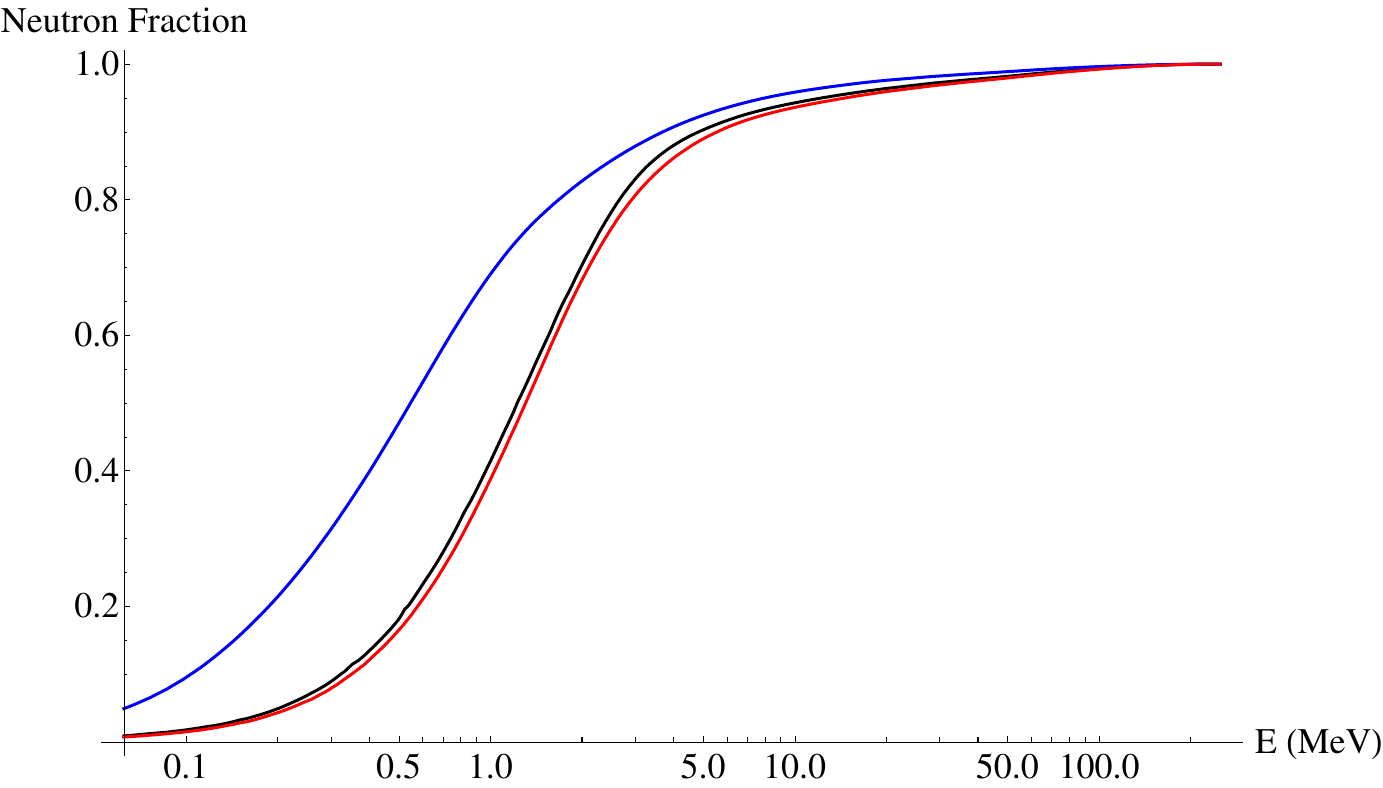}
\caption{The normalized cumulative distribution of the neutron energies upon exiting the target as produced using a 25 MeV (left panel, black), 60 MeV (left panel, blue) and 250 MeV (right panel) proton beam.  The $y$-axis is the fraction of neutrons beneath a specific energy.  On the left panel a Be target is used, while on the right panel Pb (black), W (blue) and Bi (red) targets are used.}
\label{neutprodfig}
\end{center}
\end{figure}

Including the latest data it is possible to improve the GEANT4 simulations considerably \cite{daedkam}.  However one of the main results of Sec.~\ref{transez} is that below about 25 MeV the initial neutron energy has little effect on the isotope production.  In Fig.~\ref{neutprodfig} we plot the normalized cumulative distributions of the neutron spectra  produced by 25 MeV, 60 MeV and 250 MeV proton beams, as determined by FLUKA.

As can be seen in Fig.~\ref{neutprodfig}, in each case less than 5\% of the neutrons exiting the target have energies in excess of 25 MeV and at most about 10\% have energies in excess of 10 MeV.   We have seen in Subsec.~\ref{neutsimsez} that below 10 MeV, the ${}^8$Li yield per neutron is quite independent of the neutron energy.  Therefore the shape of the neutron spectrum will have very little effect on the final ${}^8$Li yield, especially for a large target station.  This means that for our study of the effect of target station design on the ${}^8$Li yield, we only need the total normalizations of the neutron flux and not the detailed spectral shape below 10 MeV.  For example, even if the neutron energy is doubled in a 100 kg LiOD converter, the ${}^8$Li yield only falls about 5\%.  This justifies our use of unmodified GEANT4 and FLUKA in this note:  FLUKA and GEANT4 underestimate the neutron flux significantly but the missing neutrons are at energies well below the neutron multiplication threshold, where the neutron energy and so the spectral shape does not affect the ${}^8$Li yield.   It would therefore be possible to correct for this shortfall of neutrons by rescaling the ${}^8$Li yield by the ratio of neutrons observed in a fixed target experiment such as \cite{tilquin,osipenko} to those obtained by the simulation.   In Fig.~\ref{neutprodfig}, we plot the fractional distribution of neutrons and so the overall normalisation does not appear.  In the following section our results are not rescaled.

\section{The Full Simulation} \label{pienosez}

In this subsection we simulate the full experimental setup, from the proton beam to the ${}^8$Li production.  Note that the result cannot simply be obtained by folding the results of Sec.~\ref{prodsez} into Sec.~\ref{transez} because neutrons can bounce from the converter back into the target, where they may be absorbed.  This bounce-back process is only possible in a simulation which includes both the target and the converter.  In particular, we will see that bounce-back is most important for W targets, which have the highest probability of absorbing the neutrons.  On the other hand it is nearly negligible for the other targets.  The W target nonetheless is important as a granular W target is currently the favored target for the CI-ADS 250 MeV beam, even if the beam energy is increased to 600 MeV.

As FLUKA predicts lower spallation neutron yields than have been observed in experiment, one may expect that the true ${}^8$Li yields will be 20-40\% greater than those reported below in each case.

\subsection{Comparison of converters}

In this subsection we compare various converter designs.  As bounce-back results in significant neutron loss in the case of a W converter, the target has been surrounded with a 10 cm gap or vacuum sleeve in this case as described in Subsec.~\ref{manicasez}.  To increase the yield of the metallic Li converter, 5 cm of heavy water moderator has been placed between the target and the converter in this case, following the design in Ref.~\cite{daed12}.

\begin{figure} 
\begin{center}
\includegraphics[width=3.2in,height=1.5in]{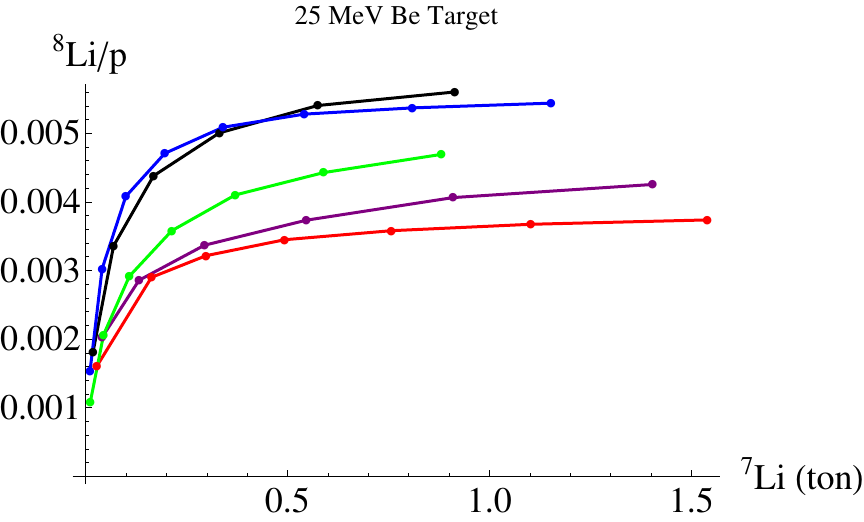}
\includegraphics[width=3.2in,height=1.5in]{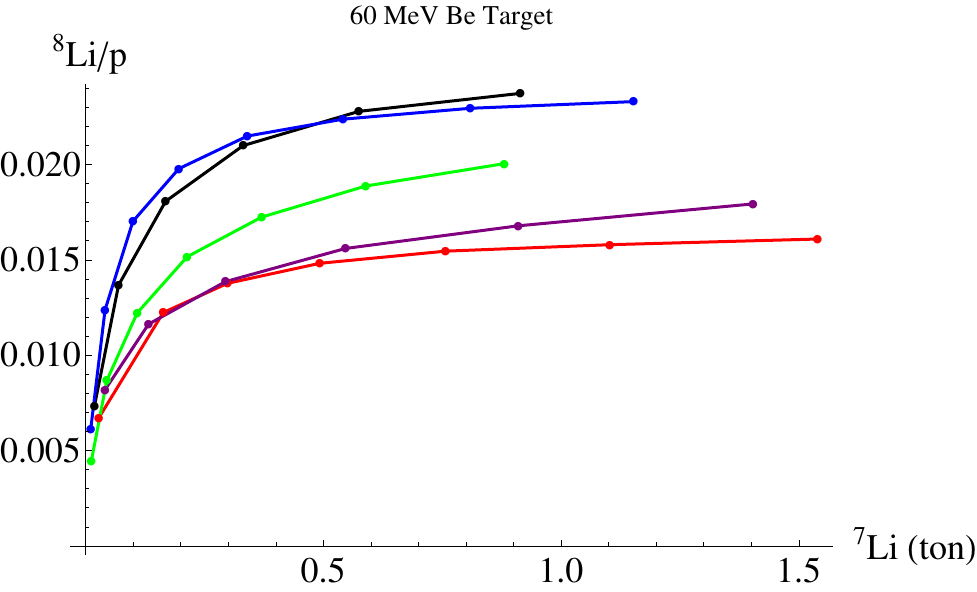}
\includegraphics[width=3.2in,height=1.5in]{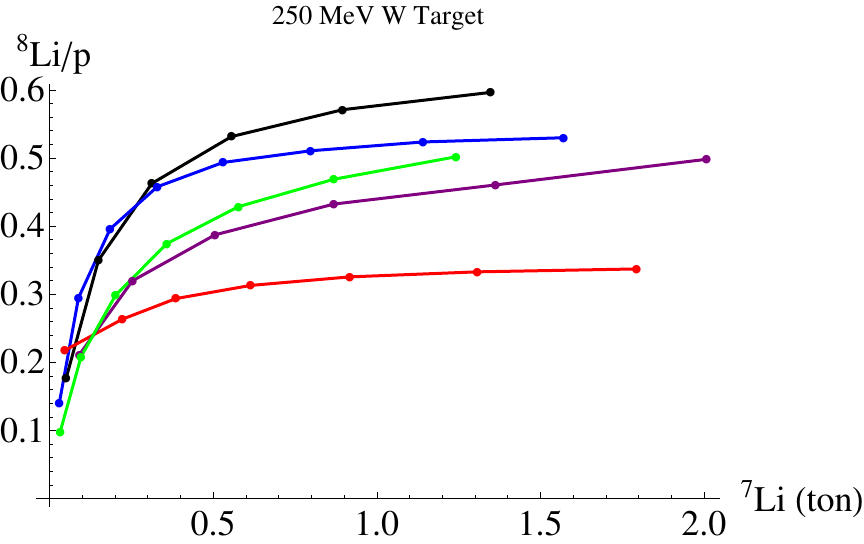}
\caption{The ${}^8$Li production given a 25 MeV, 60 MeV and 250 MeV proton beam is shown in the three panels as a function of the ${}^7$Li mass.  The target is always a cylinder of length 20 cm and radius 10 cm.  At 25 MeV and 60 MeV the target is Be.  At 250 MeV the target is W and it is surrounded by a 10 cm vacuum sleeve.  The converters are LiOD (black), LiOD$\cdot$D${}_2$O (blue), the solution (red), metallic Li (purple) and FLiBe (green).  In the case of metallic Li, a 5 cm heavy water moderator is placed inside the converter.} 
\label{totfig}
\end{center}
\end{figure}

\begin{figure} 
\begin{center}
\includegraphics[width=3.2in,height=1.5in]{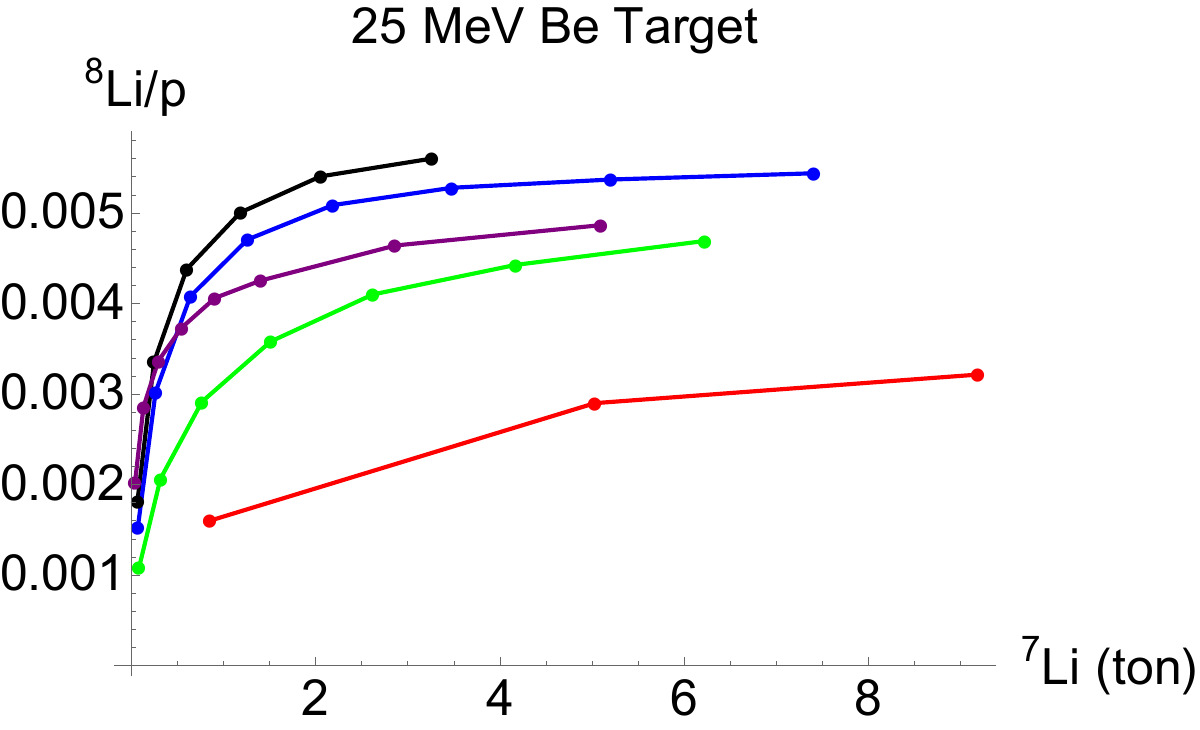}
\includegraphics[width=3.2in,height=1.5in]{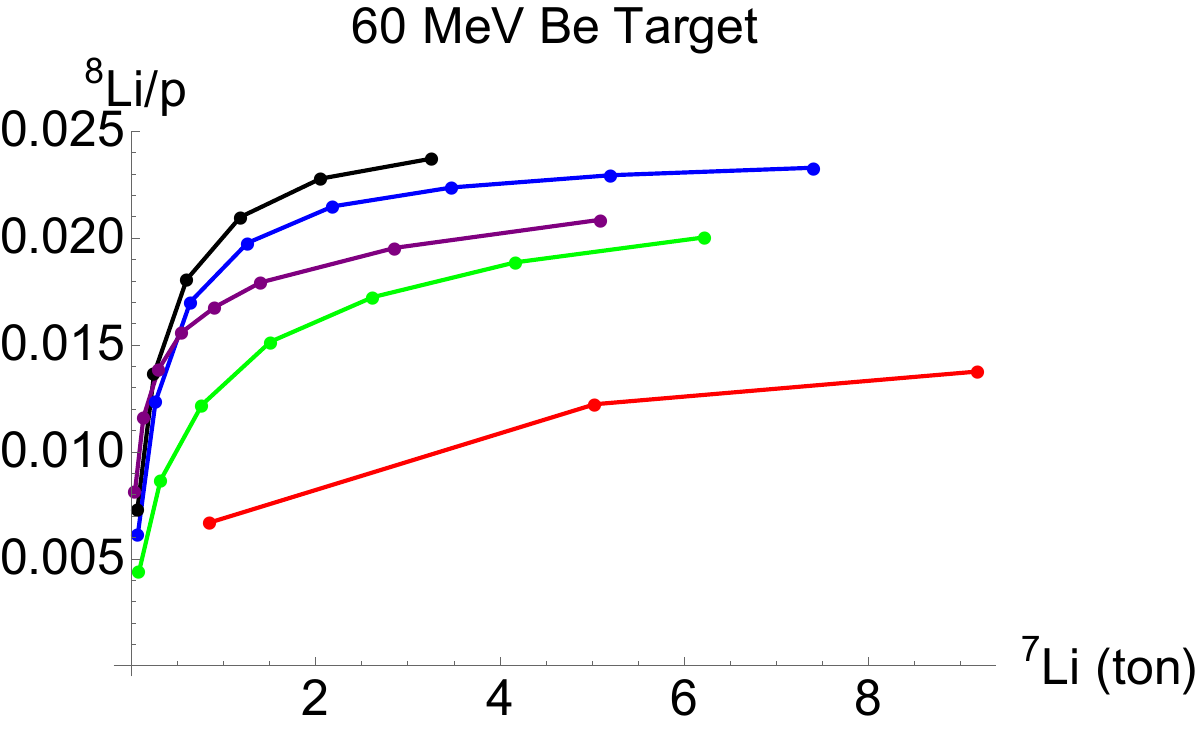}
\includegraphics[width=3.2in,height=1.5in]{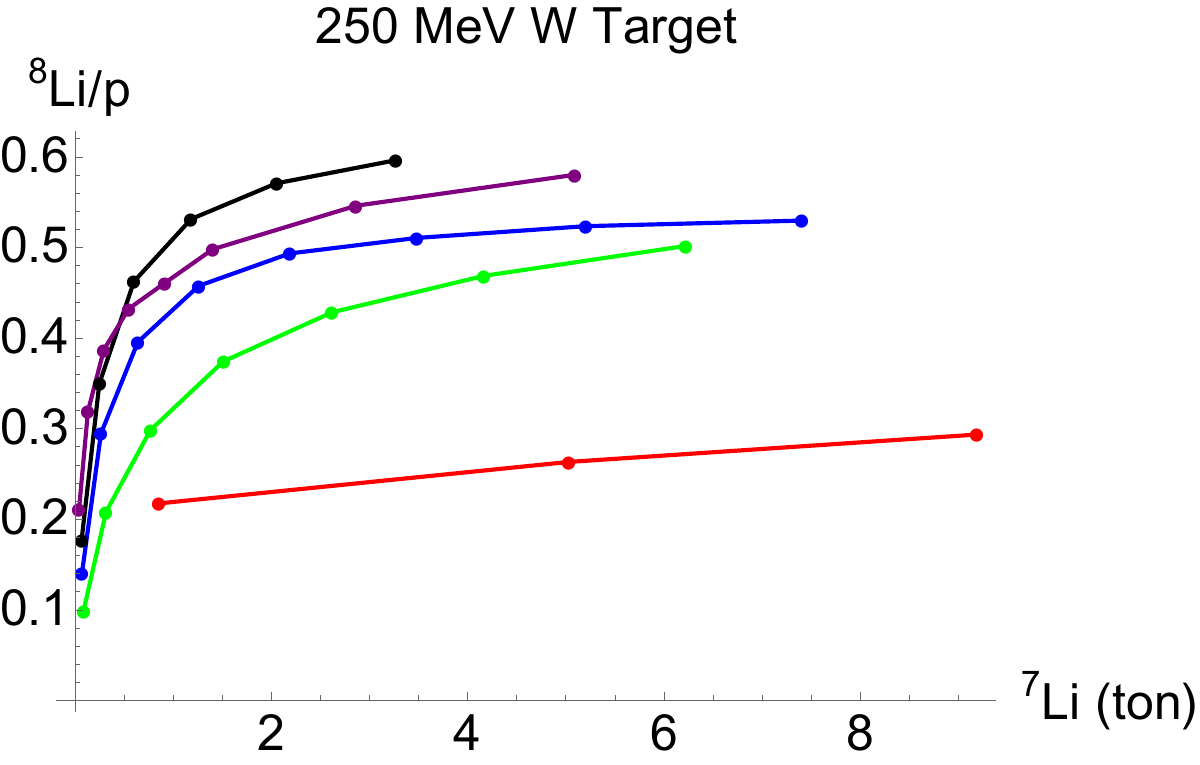}
\caption{As in Fig.~\ref{totfig} but the $x$-axis is the total converter mass instead of just the ${}^7$Li mass.} 
\label{totpesifig}
\end{center}
\end{figure}

The results are shown in Figs.~\ref{totfig} for various ${}^7$Li masses.  Our main result is apparent here, the converters in which Li is mixed with a deuterium moderator significantly outperform the others with the same total mass of ${}^7$Li, in accordance with the expectations of Ref.~\cite{russi90}.   The ${}^7$Li mass dominates the materials cost of the converter, however in Fig.~\ref{totpesifig} we have performed the same comparison fixing the {\it{total}} converter mass.  Here one finds that metallic Li is the best at very small masses.  In the case of the W target and 250 MeV beam, LiOD$\cdot$D${}_2$O suffers considerably from neutrons lost after bouncing back into the target and indeed one can see that as a result, at fixed total converter mass, it is outperformed by metallic Li.

\subsection{Liquid nitrogen cooling}

As can be seen in Table~\ref{quantab}, an important contribution to the distance that neutrons need to travel is $d_{\rm abs}$, the distance traveled between thermalization and absorption.  The ${}^8$Li to neutron ratio approaches $p_\nu$ when the size of the converter is several times $d_{\rm abs}$.  Therefore $d_{\rm abs}$ sets the scale of the converter and is responsible for the reduction in ${}^8$Li generation when the converter is smaller than that scale.  As a result, a smaller $d_{\rm abs}$ would allow for a smaller converter with the same $p_\nu$ or a larger $p_\nu$ with the same size converter.

The length scale $d_{\rm abs}$ is, according to Eqs.~(\ref{pabs}) and (\ref{dabs}), inversely proportional to the square root of the absorption cross sections $\sigma^i_{\rm abs}$.  These in turn are inversely proportional to the neutron velocity, and so to the square root of the neutron temperature.  This means that if the converter temperature is reduced by a factor of 4, by cooling with liquid nitrogen, then the neutron velocities will be reduced by a factor of 2 and so $d_{\rm abs}$ will be reduced by a factor of $\sqrt{2}$, allowing for an increased ${}^8$Li yield with a smaller and cheaper target station.

{{In liquid nitrogen one expects the absorption distance to be halved.  On the other hand, the factor of four reduction in temperature corresponds to only 4 or 5 additional D collisions, and so only about a 5\% increase in the thermalization distance.  Thus one expects the sum in quadrature of the thermalization and absorption distances to decrease if the converter is cooled to liquid nitrogen temperatures.}}

\begin{figure} 
\begin{center}
\includegraphics[width=3.2in,height=1.5in]{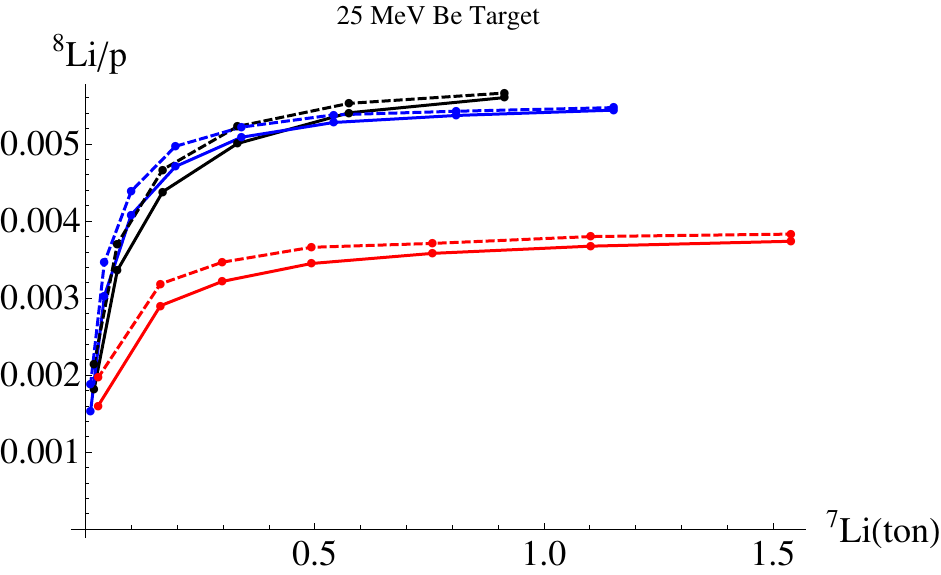}
\caption{The ${}^8$Li yield per 25 MeV proton as a function of the ${}^7$Li mass.  The black, blue and red curves correspond to LiOD, LiOD$\cdot$D${}_2$O and the solution respectively.  The solid (dashed) curves correspond to a room temperature (liquid nitrogen cooled) converter.}
\label{azotofig}
\end{center}
\end{figure}

We have rerun our simulations with the converter at liquid nitrogen temperature, again without including any cooling system in our design.  In practice the 60 MeV and 250 MeV experiments are likely to have enough money to buy large converters, and so for brevity we only report our results in the case of the 25 MeV proton beam in Fig.~\ref{azotofig}, although the cooling has a similar effect at other energies.  As expected based on the general arguments above, liquid nitrogen cooling has only a modest effect on the ${}^8$Li production.  However, in general it leads to a ${}^8$Li production with a mass $M$ of ${}^7$Li equal to that which would be obtained with a mass of about $2M$ of room temperature ${}^7$Li.  This corresponds to an improvement which is quite small for large ${}^7$Li converter masses, but approaches 20\% when the ${}^7$Li mass is less than 100 kg.

\subsection{Vacuum sleeve} \label{manicasez}

In every setup, a considerable fraction of the neutrons bounce back into the target. However only the W target has a sufficiently high neutron absorption cross section to absorb an appreciable fraction of these neutrons. Nonetheless, a W target is currently favored for the CI-ADS 250 MeV accelerator, and so this case cannot be ignored.

\begin{figure} 
\begin{center}
\includegraphics[width=3.2in,height=1.5in]{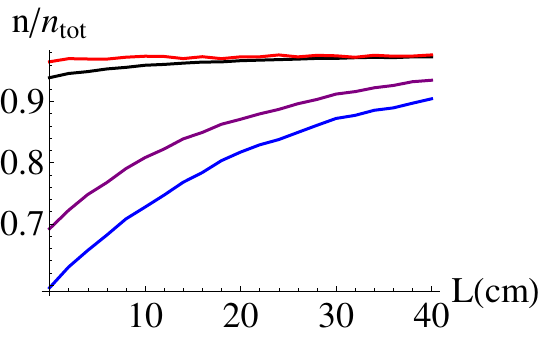}
\caption{The fraction of neutrons that are not lost to neutron bounce-back, in the case of a 250 MeV proton beam, as a function of the size of the gap between the target and the converter.  The red, black, purple and blue curves correspond to a Bi target with a LiOD$\cdot$D${}_2$O converter, a Pb target with a LiOD$\cdot$D${}_2$O converter, a W target with a LiOD converter and a W target with a LiOD$\cdot$D${}_2$O converter.}
\label{balzafig}
\end{center}
\end{figure}

If there is a gap or vacuum sleeve between the target and the converter, then some of the neutrons bouncing back from the converter will fly through the gap and reenter the converter elsewhere without ever entering the target.  As a result, a gap reduces the number of neutrons lost to bounce-back.  In Fig.~\ref{balzafig} we plot the results of FLUKA simulations of the fraction of neutrons which leave the target and are not reabsorbed in the target later, as a function of the gap size.

One can observe that in the case of the LiOD$\cdot$D${}_2$O converter, more neutrons are lost to bounce-back than in the case of the LiOD converter.  We have verified that this is a consequence of more neutrons bouncing back, and not of the energy spectrum of the bounced-back neutrons.

The main result of this study is that bounce-back can lead to a loss of as many as 40\% of the neutrons.  However, with a sufficiently large gap, this loss can be made as small as desired.  Of course, a large gap also implies that a greater quantity of ${}^7$Li is needed for the same converter thickness.  It also means that the $\overline{\nu}_e$ are created over a larger physical area, leading to a greater baseline uncertainty which reduces the sensitivity of a sterile neutrino experiment at large $\Delta$M${}^2$.

\section{Conclusions}

IsoDAR experiments provide powerful and we believe also feasible tests of sterile neutrino models that have been invoked to explain observed anomalies.  Several such experiments have been proposed at laboratories around the world.

In this note we have simulated three proposed methods for increasing the $\overline{\nu}_e$ yield of these experiments.  In the first, the ${}^7$Li converter is mixed with a D moderator.  In the second, the converter is cooled to liquid nitrogen temperatures.  In the third, a gap is placed between the target and the converter.  We have not investigated the feasibility of these modifications.  In particular, our simulations were quite idealized as we did not include a support structure for a target separated by a gap, nor cooling for the target or for the converter.  We also did not include the impurities such as K which are normally included in isotopically pure ${}^7$Li available on the market.

We have found that the utility of each of these modifications depends on the experimental setup.  For example, the purpose of the gap is to allow bounced-back neutrons to reenter the converter without passing through the target where they may be absorbed.  However, only the W target has a sufficiently high absorption cross section for bounced-back neutrons to significantly affect the $\overline{\nu}_e$ yield.  Therefore we have found that the gap is only useful in the case of the W target.  However it is quite likely that the 250 MeV CI-ADS beam will use a W target, therefore it seems likely that one will wish to incorporate this gap in the target station design for any IsoDAR experiment at that beam.

The liquid nitrogen cooling reduces the fraction of neutrons that escape the converter and the reflector to the outside.  The fraction of escapes which are prevented is significant.  However such neutron escapes are themselves only significant in the case of a thin moderator, in particular if less than 100 kg of Li is used.  Such a thin moderator would only be used either to save money, or so as to be able to obtain a higher purity at the same price.  For a very thin moderator, the increase in $\overline{\nu}_e$ yield at liquid nitrogen temperatures approaches 20\%.  However, given the hot target in the center of the target station, appreciable cooling of the converter may be impractical. In fact, it may be that the converter is appreciably above room temperature, in which case more neutrons will escape than we have simulated and so the converter will need to be larger.  We need to insure however that the converter does not become too hot, as some of these converters will thermally decompose \cite{kudo}.  In the future we intend to perform more detailed simulations, including heat dissipation, to resolve this issue.

We have found that FLiBe provides the highest ${}^8$Li and so $\overline{\nu}_e$ yield for neutron energies well above 25 MeV.  However we have also found that proton beams of energy up to 250 MeV produce negligible quantities of neutrons at such high energies.  As a result, the highest ${}^8$Li/p yields were obtained using Li compounds which include D.

Perhaps our main result is that we have confirmed the claims of Ref.~\cite{russi90} that mixing the converter with a D moderator improves the neutron capture rate appreciably.  As a result, for Li masses of order a ton or less, neutrons can thermalize anywhere in the converter volume and so far less neutrons escape, increasing the $\overline{\nu}_e$ yield considerably with respect to the metallic Li converter outside of a thin moderator proposed in Ref.~\cite{daed12}.

In this article we have determined how several modifications of the core IsoDAR target station design can potentially affect the $\overline{\nu}_e$ yield.  In the future, to drive these proposals further, we will investigate both their practicality and also their effects on the physics goals of IsoDAR experiments.  To do this, we will require simulations which correctly reproduce the shape of the neutron energy spectrum and its angular distribution and also model the target station heating and cooling.

\section* {Acknowledgement}
\noindent
We thank M. Osipenko for discussions and advice.  JE and EC are supported by NSFC grant 11375201.  EC  is also supported by the Chinese Academy of Sciences President's International Fellowship Initiative grant 2015PM063 and NSFC grant 11605247. MD is supported by the Chinese Academy of Sciences President's International Fellowship Initiative grant 2016PM043.  JE is supported by the CAS Key Research Program of Frontier Sciences grant QYZDY-SSW-SLH006.  JE, EC and MD thank the Recruitment Program of High-end Foreign Experts for support.


\end{document}

\bibitem{cads}
  Z.~Li {\it et al.},
  ``Physics design of an accelerator for an accelerator-driven subcritical system,''
  Phys.\ Rev.\ ST Accel.\ Beams {\bf 16} (2013) 8,  080101.

\bibitem{daed}
  J.~Alonso, F.~T.~Avignone, W.~A.~Barletta, R.~Barlow, H.~T.~Baumgartner, A.~Bernstein, E.~Blucher
{\it\ et al.},
  ``Expression of Interest for a Novel Search for CP Violation in the Neutrino Sector: DAE$\delta$ALUS,''
  arXiv:1006.0260 [physics.ins-det].

\bibitem{moment}
 J.~Cao {\it et al.},
  ``Muon-decay medium-baseline neutrino beam facility,''
  Phys.\ Rev.\ ST Accel.\ Beams {\bf 17} (2014) 090101
 [arXiv:1401.8125 [physics.acc-ph]].

\bibitem{spallwhite}
  M.~Elnimr {\it et al.} [OscSNS Collaboration],
  ``The OscSNS White Paper,''
  arXiv:1307.7097.

\bibitem{lund}
  E.~Baussan {\it et al.} [ESSnuSB Collaboration],
  ``A very intense neutrino super beam experiment for leptonic CP violation discovery based on the European spallation source linac,''
  Nucl.\ Phys.\ B {\bf 885} (2014) 127
  [arXiv:1309.7022 [hep-ex]].

\bibitem{laser}
S.~V.~Bulanov, T.~Esirkepov, P.~Migliozzi, F.~Pegoraro, T.~Tajima and F.~Terranova,
  ``Neutrino oscillation studies with laser-driven beam dump facilities,''
  Nucl.\ Instrum.\ Meth.\ A {\bf 540} (2005) 25
  [hep-ph/0404190].

\bibitem{isodarussi2005}
  Y.~S.~Lutostansky and V.~I.~Lyashuk,
  ``Antineutrino spectrum from a powerful reactor and neutrino converter system,''
  Phys.\ Part.\ Nucl.\ Lett.\  {\bf 2} (2005) 226
   [Pisma Fiz.\ Elem.\ Chast.\ Atom.\ Yadra {\bf 127N4} (2005) 60],
   http://ftp.jinr.ru/publish/Pepan\_letters/\\panl\_4\_2005/06\_lut.pdf .

\bibitem{isodar}
  A.~Bungau {\it et al.},
  ``Proposal for an Electron Antineutrino Disappearance Search Using High-Rate $^{8}$Li Production and Decay,''
  Phys.\ Rev.\ Lett.\  {\bf 109} (2012) 141802
  [arXiv:1205.4419 [hep-ex]].

\bibitem{fengyi}
  F.~Zhao, Y.~Li, C.~Han, Q.~Fu and X.~Chen,
  ``IsoDAR Neutrino Experiment Simulation with Proton and Deuteron Beams,''
  arXiv:1509.03922 [physics.ins-det].

\bibitem{deut62}
R.~Alba {\it et al.},
  ``Measurement of neutron yield by 62 MeV proton beam on a thick Beryllium target,''
  J.\ Phys.\ Conf.\ Ser.\  {\bf 420} (2013) 012162
  [arXiv:1208.1713 [nucl-ex]].
  
\bibitem{deut200}
N.~Pauwels {\it et al.},
 ``Experimental determination of neutron spectra produced by bombarding thick targets: Deuterons (100 MeV/u) on ${}^9$Be and ${}^{238}$U and ${}^{36}$Ar on ${}^{12}$C,''
 Nucl. Inst. and Meth. {\bf B160} (2000) 315.

\bibitem{isodarussi2015}
  Y.~S.~Lutostansky and V.~I.~Lyashuk,
  ``Intensive neutrino source on the base of lithium converter ,''
  arXiv:1503.01280 [physics.ins-det].

\bibitem{kopp}
    J.~Kopp, P.~A.~N.~Machado, M.~Maltoni and T.~Schwetz,
  ``Sterile Neutrino Oscillations: The Global Picture,''
  JHEP {\bf 1305} (2013) 050
  [arXiv:1303.3011 [hep-ph]].


\bibitem{vogel15}
 P.~Vogel, L.~Wen and C.~Zhang,
  ``Neutrino Oscillation Studies with Reactors,''
  Vogel, P., Wen, L. J. and Zhang, C., Nature Communications 6, 6935
  (2015)
  [arXiv:1503.01059 [hep-ex]].

\bibitem{lsnd}
  A.~Aguilar-Arevalo {\it et al.}  [LSND Collaboration],
  ``Evidence for neutrino oscillations from the observation of anti-neutrino(electron) appearance in a anti-neutrino(muon) beam,''
  Phys.\ Rev.\ D {\bf 64} (2001) 112007
  [hep-ex/0104049].

\bibitem{icarus}
   M.~Antonello {\it et al.} [ICARUS Collaboration],
  ``Search for anomalies in the ${\nu}_e$ appearance from a ${\nu}_{\mu}$ beam,''
  Eur.\ Phys.\ J.\ C {\bf 73} (2013) 2599
  [arXiv:1307.4699 [hep-ex]].

\bibitem{lsnd1997}
   C.~Athanassopoulos {\it et al.} [LSND Collaboration],
  ``Measurements of the reactions C-12 (electron-neutrino, e-) N-12 (g.s.) and C-12 (electron-neutrino, e-) N*-12,''
  Phys.\ Rev.\ C {\bf 55} (1997) 2078
  [nucl-ex/9705001].
  
\bibitem{lsnd2002}
 L.~B.~Auerbach {\it et al.} [LSND Collaboration],
  ``Measurements of charged current reactions of muon neutrinos on C-12,''
  Phys.\ Rev.\ C {\bf 66} (2002) 015501
  [nucl-ex/0203011].

\bibitem{juno}
  Y.~-F.~Li, J.~Cao, Y.~Wang and L.~Zhan,
  ``Unambiguous Determination of the Neutrino Mass Hierarchy Using Reactor Neutrinos,''
  Phys.\ Rev.\ D {\bf 88} (2013) 013008
  [arXiv:1303.6733 [hep-ex]].

\bibitem{noidarts}
 E.~Ciuffoli, J.~Evslin and X.~Zhang,
  ``The Leptonic CP Phase from Muon Decay at Rest with Two Detectors,''
  JHEP {\bf 1412} (2014) 051
  [arXiv:1401.3977 [hep-ph]].

\bibitem{kaoru}
  J.~Evslin, S.~F.~Ge and K.~Hagiwara,
  ``The Leptonic CP Phase from T2(H)K and Muon Decay at Rest,''
  arXiv:1506.05023 [hep-ph].

\bibitem{nova}
  C.~Backhouse,
  ``Results from MINOS and NO$\nu$A,''
  J.\ Phys.\ Conf.\ Ser.\  {\bf 598} (2015) 1,  012004
  [arXiv:1501.01016 [hep-ex]].

\bibitem{honda}
 M.~Sajjad Athar, M.~Honda, T.~Kajita, K.~Kasahara and S.~Midorikawa,
  ``Atmospheric neutrino flux at INO, South Pole and Pyh\'asalmi,''
  Phys.\ Lett.\ B {\bf 718} (2013) 1375
  [arXiv:1210.5154 [hep-ph]].

\bibitem{hondasite}
${\rm{http://www.icrr.u-tokyo.ac.jp/\tilde mhonda/nflx2014/lowe/}}$

\bibitem{genie}
C.~Andreopoulos, A.~Bell, D.~Bhattacharya, F.~Cavanna, J.~Dobson, S.~Dytman, H.~Gallagher and P.~Guzowski {\it et al.},
  ``The GENIE Neutrino Monte Carlo Generator,''
  Nucl.\ Instrum.\ Meth.\ A {\bf 614} (2010) 87
  [arXiv:0905.2517 [hep-ph]].

\bibitem{sk2011}
  K.~Bays {\it et al.}  [Super-Kamiokande Collaboration],
  ``Supernova Relic Neutrino Search at Super-Kamiokande,''
  Phys.\ Rev.\ D {\bf 85} (2012) 052007
  [arXiv:1111.5031 [hep-ex]].

\bibitem{kamlandnc}
 A.~Gando {\it et al.} [KamLAND Collaboration],
  ``A study of extraterrestrial antineutrino sources with the KamLAND detector,''
  Astrophys.\ J.\  {\bf 745} (2012) 193
  [arXiv:1105.3516 [astro-ph.HE]].

\bibitem{lenapulse}
R.~Möllenberg, F.~von Feilitzsch, D.~Hellgartner, L.~Oberauer, M.~Tippmann, V.~Zimmer, J.~Winter and M.~Wurm,
  ``Detecting the Diffuse Supernova Neutrino Background with LENA,''
  Phys.\ Rev.\ D {\bf 91} (2015) 3,  032005
  [arXiv:1409.2240 [astro-ph.IM]].

\bibitem{nova}
  R.~B.~Patterson [NOvA Collaboration],
  ``The NOvA Experiment: Status and Outlook,''
  Nucl.\ Phys.\ Proc.\ Suppl.\  {\bf 235-236} (2013) 151
  [arXiv:1209.0716 [hep-ex]].

\bibitem{dune}
  M.~Goodman,
  ``The Deep Underground Neutrino Experiment,''
  Adv.\ High Energy Phys.\  {\bf 2015} (2015) 256351.

\bibitem{hk}
 K.~Abe {\it et al.} [Hyper-Kamiokande Proto- Collaboration],
  ``Physics potential of a long-baseline neutrino oscillation experiment using a J-PARC neutrino beam and Hyper-Kamiokande,''
  PTEP {\bf 2015} (2015) 053C02
  [arXiv:1502.05199 [hep-ex]].

\end{document}